\documentclass[aps,prb,preprint,showpacs,preprintnumbers,groupedaddress]{revtex4-1}

\usepackage[final]{graphicx}
\usepackage{amsmath,amssymb,latexsym}    
\usepackage{wasysym}
\usepackage{amsfonts}
\usepackage{color}
\usepackage{nicefrac}
\usepackage[bookmarks=false,pdffitwindow=true,colorlinks,linkcolor=black,citecolor=black,filecolor=black,urlcolor=blue,pdfborder={0 0 0}]{hyperref}
\usepackage[english]{babel}

\def\Tensor#1{\overset{\text{\tiny$\bm\leftrightarrow$}}{#1}}

\begin{document}

\preprint{APS/123-QED}

\title{Effects of Crystal Anisotropy on Optical Phonon Resonances in the Mid-Infrared Second Harmonic Response of SiC}

\author{Alexander Paarmann}
\email{alexander.paarmann@fhi-berlin.mpg.de}

\author{Ilya Razdolski}

\author{Sandy Gewinner}

\author{Wieland Sch\"ollkopf}

\author{Martin Wolf}

\affiliation{Fritz-Haber-Institut der Max-Planck-Gesellschaft, Faradayweg 4-6,
14195 Berlin, Germany}

\date{\today}

\begin{abstract}
We study the effects of crystal anisotropy on optical phonon resonances in the second harmonic generation (SHG) from silicon carbide (SiC) in its Reststrahl region. By comparing experiments and simulations for isotropic 3C-SiC and anisotropic 4H-SiC in two crystal cuts, we identify several pronounced effects in the nonlinear response which arise solely from the crystal anisotropy. Specifically, we demonstrate that the axial and planar transverse optical phonon resonances selectively and exclusively appear in the corresponding tensor elements of the nonlinear susceptibility, enabling observation of an intense SHG peak originating from a weak phonon mode due to zone-folding along the c-axis of 4H-SiC. Similarly, we identify an anisotropy factor $\zeta \equiv \epsilon_\perp/\epsilon_\parallel$ responsible for a steep enhancement of the transmitted fundamental fields at the axial longitudinal optical phonon frequency, resulting in strongly enhanced SHG. We develop a general recipe to extract all these features that is directly applicable to all wurtzite-structure polar dielectrics, where a very similar behavior is expected. Our model study illustrates the opportunities for utilizing the crystal anisotropy for selectively enhancing nonlinear-optical effects in polar dielectrics, which could potentially be extended to built-in anisotropy in artificially designed hybrid materials.
\end{abstract}

\pacs{63.20.-e, 78.30.-j, 42.65.Ky}

\maketitle
The mid-infrared (mid-IR) Reststrahl spectral region of polar dielectric materials has recently attracted considerable attention due to their potential for a novel branch of nanophotonic applications based on surface phonon polaritons (SPhPs),\cite{Taubner2006,Caldwell2013,Gubbin2016, Li2016, Caldwell2014a, Dai2014} which rely on optical phonon resonanances in the dielectric response. In particular, strongly anisotropic crystals such as hexagonal boron nitride show much promise for entirely novel applications based on the natural hyperbolic character of the SPhPs.\cite{Yoxall2015,Dai2014,Caldwell2014} For weakly anisotropic crystals such as the hexagonal polytypes of silicon carbide (SiC),\cite{Nakashima1997,Mutschke1999} which has been a test ground for many recent SPhP studies,\cite{Caldwell2013,Taubner2006,Gubbin2016} the anisotropy effects are more subtle.\cite{Engelbrecht1993,Bluet1999} Additionally, the possibility to design hybrid surface polariton materials with artificially built-in anisotropy allows to tailor material properties for specific applications.\cite{Li2016,Dai2014,Caldwell2016} Considering the large class of polar dielectric materials\cite{Burfoot1979} with strongly varying degree of crystal anisotropy, novel experimental methods with particular sensitivity to the anisotropic effects in the Reststrahl spectral region are highly desirable. 

For the special case of SiC, previous studies of the interplay between the crystal anisotropy and optical phonons have employed linear optical techniques like reflectivity\cite{Engelbrecht1993, Bluet1999} and Raman spectroscopy.\cite{Nakashima1997} In general, nonlinear-optical techniques such as second harmonic generation (SHG) can provide valuable additional insights due to their intrinsic sensitivity to the crystal structure and symmetry,\cite{Shen1994, Yamada1994,Jordan1997, Niedermeier1999, Dekorsy2003,Bovino2013, Fiebig2005, Becher2015,Li2013a, Zhao2015} but have so far mostly been restricted to visible and near-IR spectral ranges. Very recently, we have introduced mid-IR SHG spectroscopy as a new experimental technique to study phonon resonances in the nonlinear-optical response,\cite{Paarmann2015} allowing to combine the sensitivity of SHG spectroscopy to the crystal symmetry with the optical phonon resonance behavior in the mid-IR. 

In this work, we experimentally study and theoretically describe the SHG response of different polytypes of SiC in the mid-infrared Reststrahl spectral region. We employ tunable narrowband mid-IR free-electron laser (FEL) pulses to generate the second harmonic signal in a reflective geometry. The SHG spectra of SiC typically exhibit two distinct resonance features, one being attributed to the zone-center transverse optical (TO) phonon resonance in the second order susceptibility $\chi^{(2)}$, and a second one in the longitudinal optical (LO) phonon range of $\sim 960-980$~cm$^{-1}$ due to resonances in the Fresnel transmission.\cite{Paarmann2015} Here we compare the spectral features and azimuthal behavior of the SHG  signals for isotropic 3C-SiC and anisotropic 4H-SiC crystals in two crystal cuts, and directly identify several pronounced effects arising solely from the crystal anisotropy. The most prominent feature is a very narrow and intense SHG resonance due to the zone-folded weak mode in 4H-SiC which is entirely absent for the isotropic polytype.


\section{Experiment}

\begin{figure}[htb]
\includegraphics[width = .6\textwidth]{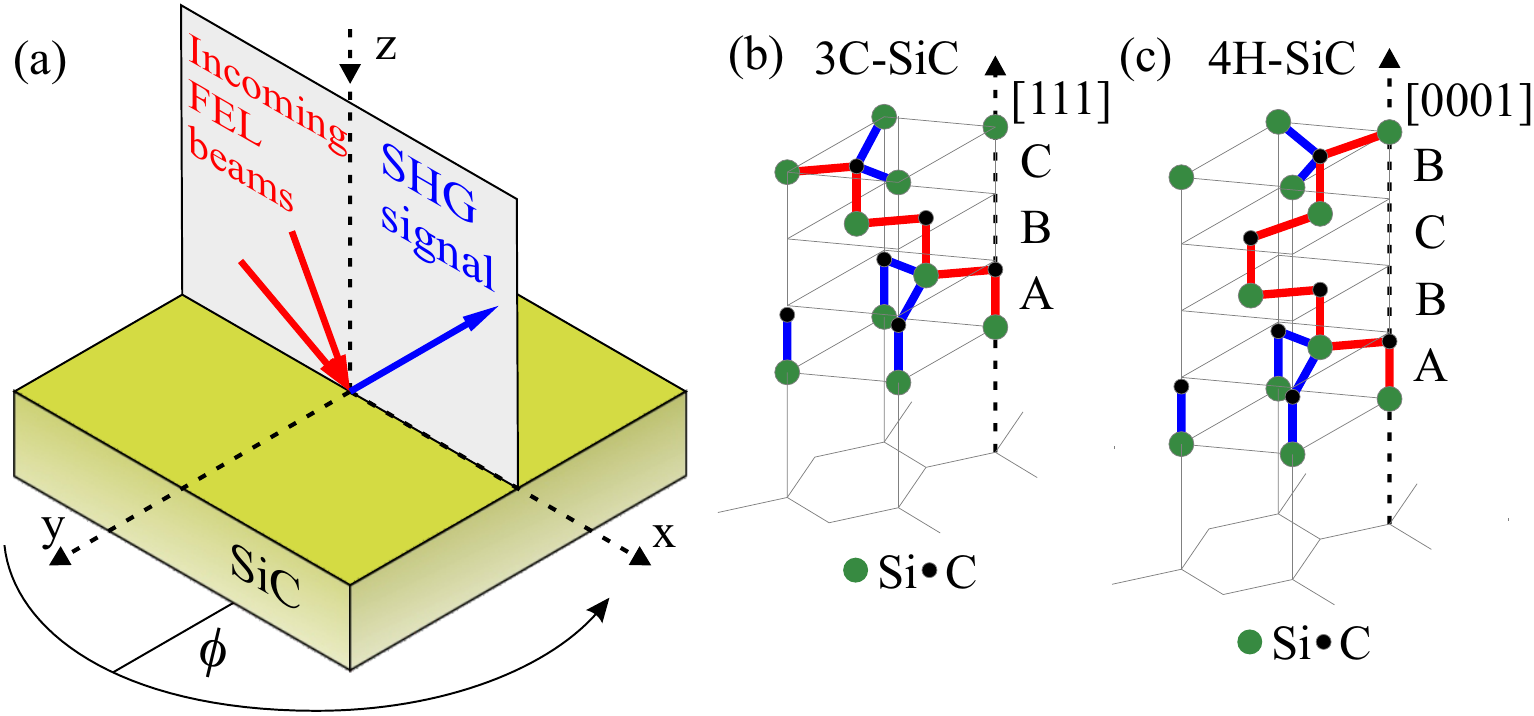}
\caption{(a) Schematic of the experimental setup and definition of the coordinate system. Non-collinear two-beam excitation with the FEL generates two-pulse correlated SHG in reflection. Rotation of the sample about the $z$-axis provides the azimuthal behavior of the SHG. (b) and (c) show schematic crystal structures of cubic 3C-SiC and hexagonal 4H-SiC, respectively.\cite{Mutschke1999}}
\label{fig:setup}
\end{figure}

The experimental setup and general approach of mid-IR SHG spectroscopy is described in detail elsewhere.\cite{Paarmann2015} In short, the incoming FEL beam is geometrically split into two equal parts that are focused to $\sim 200~\mu$m (fluence $\sim 10$~mJ/cm$^2$) and spatially overlapped on the sample at incidence angles of  $\sim 62^\circ$ and $\sim 28^\circ$, respectively, see Fig.~\ref{fig:setup}~(a). At temporal overlap of the pulses from both beams, two-pulse correlated second harmonic radiation is generated in reflection. The SHG signal is  detected with a liquid nitrogen cooled mercury cadmium telluride/indium antimony sandwich detector (Infrared Associates). Fundamental scatter contribution to the signal is minimized by short-pass spectral filtering using 5~mm thick MgF$_2$. Additionally, a $~7~\mu$m long-pass filter (LOT) is used to block intrinsic harmonics generated by the FEL. Scanning the FEL wavelength using the undulator gap results in spectroscopic measurement of the SHG signal. Sample rotation about the surface normal by an angle $\phi$ at fixed FEL wavelength is used to measure the azimuthal behavior of the SHG, see Fig.~\ref{fig:setup}~(a).  

Details on the IR-FEL are given elsewhere\cite{Schollkopf2015}. In short, the electron gun is operated at a micropulse repetition rate of 1~GHz with an electron macropulse duration of 10~$\mu$s and a  macropulse repetition rate of 10~Hz. The electron energy is set to 31~MeV, allowing to tune the FEL output wavelength between $\sim7-18$~$\mu$m ($\sim1400-550$~cm$^{-1}$) using the motorized undulator gap. The cavity desynchronism is set to $\Delta L = 2\lambda$, causing narrowband operation\cite{Schollkopf2015} with typical full-width-at-half-maximum (FWHM) of $\sim$~5~cm$^{-1}$. The FEL beam is linearely polarized; polarization rotation by 90$^\circ$ for one or both beams is realized by two subsequent wire-grid polarizers (Thorlabs), set to 45$^\circ$ and 90$^\circ$ with respect to the incoming polarization, respectively. A third polarizer is used to select a specific polarization component of the SHG signal. 

The samples investigated here are semi-insulating single crystals of SiC. To study the effect of anisotropy, we use three different samples: (i) cubic 3C-SiC with the [100] crystal axis parallel to the $z$-axis, i.e., the surface normal, (ii) c-cut 4H-SiC with the [0001] hexagonal crystal axis, which is also the optical axis of the uniaxial crystal, parallel to $z$, and (iii) a-cut 4H-SiC with the optical axis in the  $x-y$ sample surface plane, see Fig.~\ref{fig:setup}.


\section{Theory}

\subsection{Anisotropy in the mid-IR linear optical response}
The effects of crystal anisotropy on the linear optical response of hexagonal SiC polytypes was discussed before.\cite{Engelbrecht1993,Nakashima1997,Bluet1999} In uniaxial media, two different solutions of the wave equation exist, describing ordinary and extraordinary waves. In consequence, the resonance frequencies of the lattice, TO and LO phonon modes, depend on the propagation direction. For SiC, the resulting frequency splittings are small ($\sim$~10~cm$^{-1}$) as compared to the TO-LO splitting ($\sim$~160~cm$^{-1}$). 

Additionally, the stacking of different atomic layers along the $c$-axis, see Fig.~\ref{fig:setup} (c), in the hexagonal crystal structure acts as a natural super-lattice, resulting in weakly IR-active modes originating from the zone-folded LO branch.\cite{Nakashima1997,Bluet1999} For 4H-SiC, one such zone-folded weak mode can be observed at 838~cm$^{-1}$ corresponding to a reduced wavevector $x=q/q_{max}=1$,\cite{Bluet1999} with $q_{max}=\pi/c$ and $c$ the lattice constant along the hexagonal $[0001]$ direction.

The linear optical response of insulating 4H-SiC in the optical phonon spectral region is fully described by a classical dielectric function model, explicitly accounting for the crystal anisotropy effects:\cite{Bluet1999}
\begin{eqnarray}
\label{eq:eps_perp}
\epsilon_\perp(\omega) &=& \epsilon^\infty_\perp \Big ( 1 + \frac{\omega_{LO,\perp}^2 - \omega_{TO,\perp}^2 }{\omega_{TO,\perp}^2 - \omega^2 - i\omega\gamma_{TO,\perp}} \Big ) \\
\epsilon_\parallel(\omega) &=& \epsilon^\infty_\parallel \Big ( 1 + \sum_{j=0,1} \frac{\omega_{LO,\parallel j}^2 - \omega_{TO,\parallel j}^2}{\omega_{TO,\parallel j}^2 - \omega^2 - i\omega\gamma_{TO,\parallel j}}\Big ),
\label{eq:eps_par}
\end{eqnarray}
where  $\epsilon_{\perp(\parallel)}$ is the dielectric function for electric field vectors perpendicular (parallel) to the $c$-axis, and $\omega_{TO,\perp(\parallel)}$ the respective planar (axial) phonon modes.\cite{Nakashima1997} Further, $\omega_{TO,\parallel 0} \equiv \omega_{TO,\parallel}$  corresponds to the strong axial optical phonon ($x=0$), and $\omega_{TO,\parallel 1} \equiv \omega_{zf}$ to the zone-folded weak mode ($x=1$). For cubic 3C-SiC, the dielectric response is isotropic and takes the form of Eq.~\ref{eq:eps_perp}. All phonon frequencies and damping constants $\gamma$ used for the calculations in this work are summarized in Tab.~\ref{tab:freq}.

\begin{table}[bh]
\begin{tabular}{|cl|c|c|c|c|}
\hline
Polytype & & $\omega_{TO}$  & $\gamma_{TO}$  & $\omega_{LO}$  &  $\epsilon_\infty$ \\
 & & (cm$^{-1}$)  & (cm$^{-1}$)  & (cm$^{-1}$)  &   \\
\hline
3C & & 797.0 & 4 & 977.3 & 6.49 \\
\hline
4H & planar strong  & 796.6 & 2 & 972.7 & 6.56 \\
     &  axial  strong & 783.6 & 2 & 967.7 & 6.78 \\
     & axial weak   & 838.0 & 1.3 & 838.9 &  \\
\hline
\end{tabular}
\caption{Optical phonon frequencies $\omega$ and damping constants $\gamma$ (both in cm$^{-1}$),\cite{Mutschke1999,Nakashima1997,Bluet1999} as well as high frequency dielectric constants\cite{Mutschke1999} $\epsilon_\infty$  of 3C-SiC and 4H-SiC used for fitting the experimental data throughout this work.}
\label{tab:freq}
\end{table}

\subsection{Anisotropy in the SHG}

The theory of mid-infrared SHG in the Reststrahl region with optical phonon resonances was introduced in a previous work.\cite{Paarmann2015} Here we extend this theory to specifically discuss the effects of crystal anisotropy on the Reststrahl SHG response. We start with a general description before developing the theory for each of the three samples considered in this work.

Two tunable IR beams of frequency $\omega$ with incoming wave vectors $\vec{k}^{\mathrm{air}}_{1}$ and $\vec{k}^{\mathrm{air}}_{2}$ at angles of incidence $\theta^i_1$ and $\theta^i_2$, respectively, impinge on the sample. Second harmonic radiation is generated in reflection at frequency $\omega_{SHG} = 2\omega$ at an angle $\theta^r_{SHG} = \arcsin[(\sin\theta^i _1 + \sin\theta^i _1)/2]$ , see Fig.~\ref{fig:setup} (a). The second order nonlinear polarization inside the crystal in general takes the form
\begin{equation}
\vec{P}^{NL}(2\omega) \propto \Tensor\chi^{(2)}(-2\omega;\omega,\omega):(\Tensor L_1(\omega) \vec E_1(\omega) ) (\Tensor L_2(\omega) \vec E_2(\omega) ),
\label{eq:PSHG}
\end{equation}
where $\Tensor L_{1(2)}$ are the Fresnel transmission tensors for the two incident beams and $\vec E_{1(2)}$ are the incident electric field vectors. For crystals lacking inversion symmetry, the surface contribution to the second order nonlinear signals is typically negligible,\cite{Shen1989} resulting in symmetry properties of the second order nonlinear susceptibility tensor $\Tensor\chi^{(2)}$ given by the bulk crystal symmetry. The total reflected second harmonic intensity arises from projecting the nonlinear polarization onto the field direction of the reflected SH beam after considering the Fresnel transmission of the nonlinear polarization components back into air: 
\begin{eqnarray}
I(2\omega)  \propto \big |( \Tensor L_{SHG}(2\omega) \vec{P}^{NL}(2\omega)) \cdot \vec e_{SHG}\big | / \Delta k^2,
\label{eq:ISHG}
\end{eqnarray}
where $\vec e_{SHG}$ defines the field direction of the SHG beam, $\Tensor L_{SHG}$ the Fresnel tensor for the reflected SHG. Further, $\Delta k^2 = |\vec k^{\mathrm{SiC}}_{SHG} - \vec{k}^{\mathrm{SiC}}_{1} - \vec{k}^{\mathrm{SiC}}_{2}|^2$ accounts for the wave vector mismatch in reflection, with $\vec k^{\mathrm{SiC}}_{1(2),SHG}$ the first (second) transmitted fundamental and generated SHG wave vectors inside the crystal, respectively. We note that  $\vec e_{SHG}$ is defined by the SHG polarization component, i.e., $\vec e_{SHG} = (\sin\theta^r_{SHG}, 0, - \cos\theta^r_{SHG})$ for P-polarized detection and $\vec e_{SHG} = (0, 1, 0)$ for S-polarized detection.

The theoretical description of different sample orientations and azimuthal behavior requires transformation of the $\chi^{(2)}$ tensor from the crystal frame into the laboratory frame. The $\chi^{(2)}$ elements expressed in the laboratory frame coordinates $(x,y,z)$  (Fig.~\ref{fig:setup}) can in general be derived from the known $\chi^{(2)}$ elements in the crystal coordinates $(a,b,c)$  using\cite{Shen2003}
\begin{eqnarray}
\chi^{(2)}_{ijk} = \sum_{lmn} \chi^{(2)}_{lmn} (\hat i \cdot \hat l)(\hat j \cdot \hat m)(\hat k \cdot \hat n),
\label{eq:trafo}
\end{eqnarray}
where $(\hat i, \hat j,\hat k)$ and $(\hat l, \hat m,\hat n)$ are the basis vectors of the laboratory and crystal frame, respectively. 

The theory of the $\chi^{(2)}$  dispersion involving optical phonon resonances was derived by Flytzanis.\cite{Flytzanis1972} He showed that, for zincblende type crystals, three different resonant contributions with amplitudes $C_{1,2,3}$ need to be considered.  The parameters $C_{1,2,3}$ can be related to microscopic properties of the material and are denoted Faust-Henry coefficient,\cite{Faust1966} electrical and mechanical anharmonicity, respectively.\cite{Flytzanis1972,Roman2006} Flytzanis and later ab-initio work by Roman et al.\cite{Roman2006} only considered III-V zincblende semiconductors which have a single unique non-zero component of the $\chi^{(2)}$ tensor. Therefore, this theory, while directly applicable to 3C-SiC, has been extended here to account for the anisotropy in hexagonal SiC:
\begin{widetext}
\begin{eqnarray} \nonumber
\chi^{(2)}_{ijk}(-2\omega;\omega,\omega) & = \chi^{(2)}_{\infty,ijk} \Big [ 1 & +\,  C_{1,ijk} \left( \frac{1}{D_i(2\omega)} + \frac{1}{D_j(\omega)} + \frac{1}{D_k(\omega)} \right ) +
\\ \nonumber
& & + \,C_{2,ijk} \left ( \frac{1}{D_i(2\omega)D_j(\omega)} + \frac{1}{D_i(2\omega)D_k(\omega)} + \frac{1}{D_j(\omega)D_k(\omega)} \right ) 
\\
& & + \,C_{3,ijk} \left ( \frac{1}{D_i(2\omega)D_j(\omega)D_k(\omega) } \right ) \Big ] 
\label{eq:chi2}
\end{eqnarray}  
\end{widetext}
with $\chi^{(2)}_{\infty,ijk}$ the non-resonant, high frequency second order electronic susceptibility and $D_i(\omega) = 1 - \omega^2/\omega^2_{TO,i} - i\gamma_i\omega/\omega_{TO,i}^2$. Here, $\omega_{TO,i}=\omega_{TO, \perp(\parallel)}$ is the resonant planar (axial) TO phonon frequency with damping $\gamma_i$ for  electric fields along the $i$-axis of the crystal. For tensor components involving the axial modes along the  $c$-axis, additional contributions arise for the weak zone-folded mode, whereby here we neglect cross terms of the strong and weak axial phonons in the $C_2$ and $C_3$ expressions.


\subsection{Isotropic 3C-SiC}

For cubic 3C-SiC, the Fresnel tensor is diagonal and its components are straightforwardly derived from Maxwell's equations. The non-zero components are

\begin{eqnarray} \nonumber
L_{xx}(\omega, \theta^i) &=& \frac{2 k^{\mathrm{SiC}}_{z}(\omega, \theta^i)}{\epsilon(\omega) k^{\mathrm{air}}_z(\omega, \theta^i) + k^{\mathrm{SiC}}_{z}(\omega, \theta^i)}
\\ \nonumber
L_{yy}(\omega, \theta^i) &=& \frac{2 k^{\mathrm{air}}_{z}(\omega, \theta^i)}{k^{\mathrm{air}}_z(\omega, \theta^i) + k^{\mathrm{SiC}}_{z}(\omega, \theta^i)}
\\
L_{zz}(\omega, \theta^i) &=& \frac{2 k^{\mathrm{air}}_{z}(\omega, \theta^i)}{\epsilon(\omega) k^{\mathrm{air}}_z(\omega, \theta^i) + k^{\mathrm{SiC}}_{z}(\omega, \theta^i)}
\label{eq:fresnel.3C}
\end{eqnarray}
with $\epsilon(\omega)$ the only unique element of the diagonal dielectric tensor, $k^{air}_z = \frac{2\pi\omega}{c_0} \cos\theta^i$ and $k^{\mathrm{SiC}}_{z}(\omega,\theta^{i})$ the $z$-component of the complex wave vector inside the medium:\cite{Mosteller1968}
\begin{eqnarray}
k^{\mathrm{SiC}}_{z}(\omega, \theta^i) &=& \frac{2\pi\omega}{c_0}\sqrt{\epsilon(\omega) - \sin^2\theta^i},
\label{eq:k2z.3C}
\end{eqnarray} 
where $c_0$ is the speed of light in vacuum.

The $\chi^{(2)}$ tensor for zincblende type crystals ($\bar 43m$ point group symmetry) contains only a single unique non-zero element. The non-zero components are:\cite{Shen2003,Yamada1994,,Jordan1997}
\begin{equation}
\chi^{(2)}_{abc} = \chi^{(2)}_{acb} = \chi^{(2)}_{bca} = \chi^{(2)}_{bac} = \chi^{(2)}_{cba} = \chi^{(2)}_{cab}.
\end{equation}
Applying the crystal to laboratory frame transformation, Eq.~\ref{eq:trafo}, reveals the azimuthal behavior to exhibit a four-fold symmetry for all polarization geometries  with non-zero SHG signals,\cite{Yamada1994,Jordan1997} i.e., SPP (denoting S-polarized SHG and P-polarization for both incoming beams), PPP, PSS, PSP, SSP, PPS, and SPS;  only for SSS no SHG signals are expected.  As an example, the expressions for the SHG intensity for the PPP, PSS, and SPP as a function of frequency and azimuthal angle $\phi$ have the following form:
\begin{widetext}
\begin{eqnarray} \nonumber
I_{PPP} (2\omega, \phi) &\propto& \big | \big [ \cos\theta^r_{SHG}  L_{xx} (2\omega,\theta^r_{SHG}) \big (
L_{xx}(\omega,\theta^i_1)L_{zz}(\omega,\theta^i_2) + L_{zz}(\omega,\theta^i_1)L_{xx}(\omega,\theta^i_2) \big )   +  \\ \nonumber
& & + \sin\theta^r_{SHG} L_{zz} (2\omega,\theta^r_{SHG})  L_{xx}(\omega,\theta^i_1)L_{xx}(\omega,\theta^i_2)  \big ]
 \times  \chi^{(2)}_{abc}(-2\omega;\omega,\omega) \sin(2\phi)  \big |^2/\Delta k^2 , \\ \nonumber
I_{PSS} (2\omega, \phi) &\propto& \big | L_{zz} (2\omega,\theta^r_{SHG}) \big [ L_{yy}(\omega,\theta^i_1)L_{yy}(\omega,\theta^i_2) \big ] \times  \chi^{(2)}_{abc}(-2\omega;\omega,\omega) \sin(2\phi)  \big |^2/\Delta k^2 , \\  \nonumber
I_{SPP} (2\omega, \phi) &\propto& \big | L_{yy} (2\omega,\theta^r_{SHG}) \big [ L_{xx}(\omega,\theta^i_1)L_{zz}(\omega,\theta^i_2) + L_{zz}(\omega,\theta^i_1)L_{xx}(\omega,\theta^i_2)\big ] \times  \\
& & \times \chi^{(2)}_{abc}(-2\omega;\omega,\omega) \cos(2\phi)  \big |^2/\Delta k^2.  
\label{eq:SHG.3C}
\end{eqnarray}
\end{widetext}


\subsection{Anisotropic 4H-SiC: c-cut ($\mathbf{c\parallel z}$)}

For c-cut 4H-SiC, the dielectric tensor is still diagonal, however now with $\epsilon_{xx}=\epsilon_{yy}=\epsilon_\perp \neq \epsilon_{zz}=\epsilon_\parallel$. Since $\epsilon_{xx}=\epsilon_{yy}$, no azimuthal dependence of the linear optical quantities is expected. However, the effects of the crystal anisotropy enter into the Fresnel coefficients since $\epsilon_{xx(yy)} \neq \epsilon_{zz}$, resulting in:\cite{Paarmann2015}

\begin{eqnarray} \nonumber
L_{xx}(\omega, \theta^i) &=& \frac{2 k^{\mathrm{SiC}}_{z,e}(\omega, \theta^i)}{\epsilon_\perp(\omega) k^{\mathrm{air}}_z(\omega, \theta^i) + k^{\mathrm{SiC}}_{z,e}(\omega, \theta^i)}
\\ \nonumber
L_{yy}(\omega, \theta^i) &=& \frac{2 k^{\mathrm{air}}_{z}(\omega, \theta^i)}{k^{\mathrm{air}}_z(\omega, \theta^i) + k^{\mathrm{SiC}}_{z,o}(\omega, \theta^i)}
\\
L_{zz}(\omega, \theta^i) &=& \frac{\epsilon_\perp(\omega)}{\epsilon_\parallel(\omega)}\frac{2 k^{\mathrm{air}}_{z}(\omega, \theta^i)}{\epsilon_\perp(\omega) k^{\mathrm{air}}_z(\omega, \theta^i) + k^{\mathrm{SiC}}_{z,e}(\omega, \theta^i)},
\label{eq:fresnel.4Hc}
\end{eqnarray}
with $k^{\mathrm{SiC}}_{z,o(e)}(\omega)$ the $z$-component of the wave vectors of ordinary (extraordinary) waves inside the crystal:\cite{Mosteller1968}
\begin{eqnarray} \nonumber
k^{\mathrm{SiC}}_{z,o}(\omega, \theta^i) &=& \frac{2\pi\omega}{c_0}\sqrt{\epsilon_\perp(\omega) - \sin^2(\theta^i)}
\\ 
k^{\mathrm{SiC}}_{z,e}(\omega, \theta^i) &=& \frac{2\pi\omega}{c_0}\sqrt{\epsilon_\perp(\omega) - \frac{\epsilon_\perp(\omega)}{\epsilon_\parallel(\omega)} \sin^2(\theta^i)}.
\label{eq:k2z}
\end{eqnarray}

A noteworthy feature in Eq.~\ref{eq:fresnel.4Hc} is the anisotropy factor $\zeta \equiv \epsilon_\perp/\epsilon_\parallel$ appearing in $L_{zz}$, which diverges at the zero-crossing of $\epsilon_\parallel$, i.e., at the axial LO phonon frequency. In consequence, one important effect of the crystal anisotropy is a steep enhancement of the transmitted field amplitudes at this frequency.

4H-SiC has a hexagonal structure of point-group symmetry $6mm$, leading to three unique non-zero elements of the $\chi^{(2)}$ tensor and the following non-zero components:\cite{Shen2003}
\begin{eqnarray}
\chi^{(2)}_{ccc}, \chi^{(2)}_{caa} = \chi^{(2)}_{cbb}, \chi^{(2)}_{aca} = \chi^{(2)}_{bcb} = \chi^{(2)}_{aac} = \chi^{(2)}_{bbc}. 
\label{eq:chi2_elements}
\end{eqnarray} 
Due to the full symmetry between the $a$ and $b$-axes in Eq.~\ref{eq:chi2_elements}, no azimuthal dependence of the SHG signal is expected for c-cut 4H-SiC where the crystal to laboratory transformation is trivial. Non-zero SHG signals are expected for PSS, PPP, SPS, and SSP geometries. We here show the expressions for the SHG intensities for PSS and SPS:
\begin{eqnarray} \nonumber
I_{PSS} &\propto& \big | L_{zz}(2\omega,\theta^r_{SHG})  \sin\theta^r_{SHG} \chi^{(2)}_{caa}(-2\omega;\omega,\omega) \times \\ \nonumber
& &  \times L_{yy}(\omega,\theta^i_1)L_{yy}(\omega,\theta^i_2)  \big |^2/\Delta k_{PSS}^2 \\ \nonumber
I_{SPS} &\propto& \big | L_{yy}(2\omega,\theta^r_{SHG})  \chi^{(2)}_{aca}(-2\omega;\omega,\omega) \times \\
& &  \sin \theta^i_1 \times L_{zz}(\omega,\theta^i_1) L_{yy}(\omega,\theta^i_2)  \big |^2/\Delta k_{SPS}^2. 
\label{eq:SHG.4Hc}
\end{eqnarray}

The respective expression for SSP geometry has a similar form, while PPP is considerably more complex with all three unique $\chi^{(2)}$ components entering into the formula. In Eq.~\ref{eq:SHG.4Hc}, the indices of the wave vector mismatch $\Delta k_{PSS}$ and $\Delta k_{SPS}$ account for ordinary and extraordinary wave vectors of the respective fundamental and SHG waves inside the crystal. 


\subsection{Anisotropic 4H-SiC: a-cut ($\mathbf{c\perp z}$)}

For the optical axis lying in the surface plane as in a-cut 4H-SiC, the azimuthal behavior of both linear and nonlinear quantities is considerably more complex than for the cases discussed above. The dielectric and the Fresnel transmission tensor are then diagonal only for the $c$-axis parallel to $x$ or $y$ in Fig.~\ref{fig:setup}~(a). In general, both, ordinary and extraordinary, rays will be excited in the sample simultaneously, leading to four combinations of beam pairs (ordinary and extraordinary from beam 1 and beam 2, respectively)  generating second harmonic polarizations that interfere upon emission of the SHG wave. In consequence, the number of terms contributing to the SHG intensity, Eq.~\ref{eq:ISHG}, is too large to be written down in a simple form comparable to Eqs.~\ref{eq:SHG.3C} or \ref{eq:SHG.4Hc}. The general form of the azimuthal and spectral dependence of the nonlinear polarization is given by the sum over all contributions:
\begin{eqnarray} \nonumber
P^{SHG}_i(\omega,\phi)& \propto &\sum_{jkj'k'=x,y,z}  \sum_{r_1,r_2=o,e} \chi_{ijk}^{(2)}(-2\omega;\omega\omega,\phi) \times \\
& & \times L_{jj',r_1} (\omega,\theta^i_1,\phi)L_{kk',r_2}(\omega,\theta^i_2,\phi).
\label{eq:SHG.4Ha.general}
\end{eqnarray}

In the following, we restrict the formalism to the special case of $c\parallel y$. Nonetheless, we also calculate the full azimuthal dependence of the SHG signals using Eq.~\ref{eq:SHG.4Ha.general} by
employing the general Fresnel transmission formalism derived by Lekner\cite{Lekner1991} and the general form of the crystal to laboratory transformation of the $\chi^{(2)}$ tensor, Eq.~\ref{eq:trafo}.

For the special case of $c\parallel y$, the Fresnel coefficients take a form similar to Eqs~\ref{eq:fresnel.3C} and \ref{eq:fresnel.4Hc}: 
\begin{eqnarray} \nonumber
L_{xx}(\omega, \theta^i) &=& \frac{2 k^{\mathrm{SiC}}_{z,o}(\omega, \theta^i)}{\epsilon_\perp(\omega) k^{\mathrm{air}}_z(\omega, \theta^i) + k^{\mathrm{SiC}}_{z,o}(\omega, \theta^i)}
\\ \nonumber
L_{yy}(\omega, \theta^i) &=& \frac{2 k^{\mathrm{air}}_{z}(\omega, \theta^i)}{k^{\mathrm{air}}_z(\omega, \theta^i) + k^{\mathrm{SiC}}_{z,e}(\omega, \theta^i)}
\\
L_{zz}(\omega, \theta^i) &=& \frac{2 k^{\mathrm{air}}_{z}(\omega, \theta^i)}{\epsilon_\perp(\omega) k^{\mathrm{air}}_z(\omega, \theta^i) + k^{\mathrm{SiC}}_{z,o}(\omega, \theta^i)}
\label{eq:fresnel.4Ha}
\end{eqnarray}
where  $k^{\mathrm{SiC}}_{z,o(e)}(\omega, \theta^i)$ where defined in Eq.~\ref{eq:k2z}. 

For $c\parallel y$, Eq.~\ref{eq:trafo}  yields the following non-zero components of the laboratory frame $\chi^{(2)}$ tensor:
\begin{eqnarray} \nonumber
\chi^{(2)}_{xyx} = \chi^{(2)}_{xxy} = \chi^{(2)}_{zyz} = \chi^{(2)}_{zzy} & = & \chi^{(2)}_{aca} \\ \nonumber
\chi^{(2)}_{yxx} = \chi^{(2)}_{yzz} & =  &\chi^{(2)}_{caa} \\
\chi^{(2)}_{yyy} & = & \chi^{(2)}_{ccc}
\end{eqnarray}
leading to non-zero SHG signal expected for SPP, SSS, PSP, and PPS polarization geometries (exactly complementary to the c-cut geometries with non-zero SHG signals). As an example, the SHG intensities for SPP and SSS are
\begin{widetext}
\begin{eqnarray}
\nonumber
I_{SPP} (2\omega, \phi=\pi/2) &\propto& \big | L_{yy} (2\omega,\theta^r_{SHG}) \big [L_{xx}(\omega,\theta^i_1)L_{xx}(\omega,\theta^i_2) + L_{zz}(\omega,\theta^i_1)L_{zz}(\omega,\theta^i_2) \big ] \times  \\ \nonumber
& & \times \chi^{(2)}_{caa}(-2\omega;\omega,\omega)  \big |^2/\Delta k_{SPP}^2,  \\ 
I_{SSS} (2\omega, \phi=\pi/2) &\propto& \big | L_{yy} (2\omega,\theta^r_{SHG}) L_{yy}(\omega,\theta^i_1)L_{yy}(\omega,\theta^i_2)  \times  \chi^{(2)}_{ccc}(-2\omega;\omega,\omega)  \big |^2/\Delta k_{SSS}^2.
\label{eq:SHG.4Ha}
\end{eqnarray}
\end{widetext}


\section{Experimental Results}


\subsection{Isotropic 3C-SiC}

\begin{figure}[thb]
\includegraphics[width=.6\textwidth]{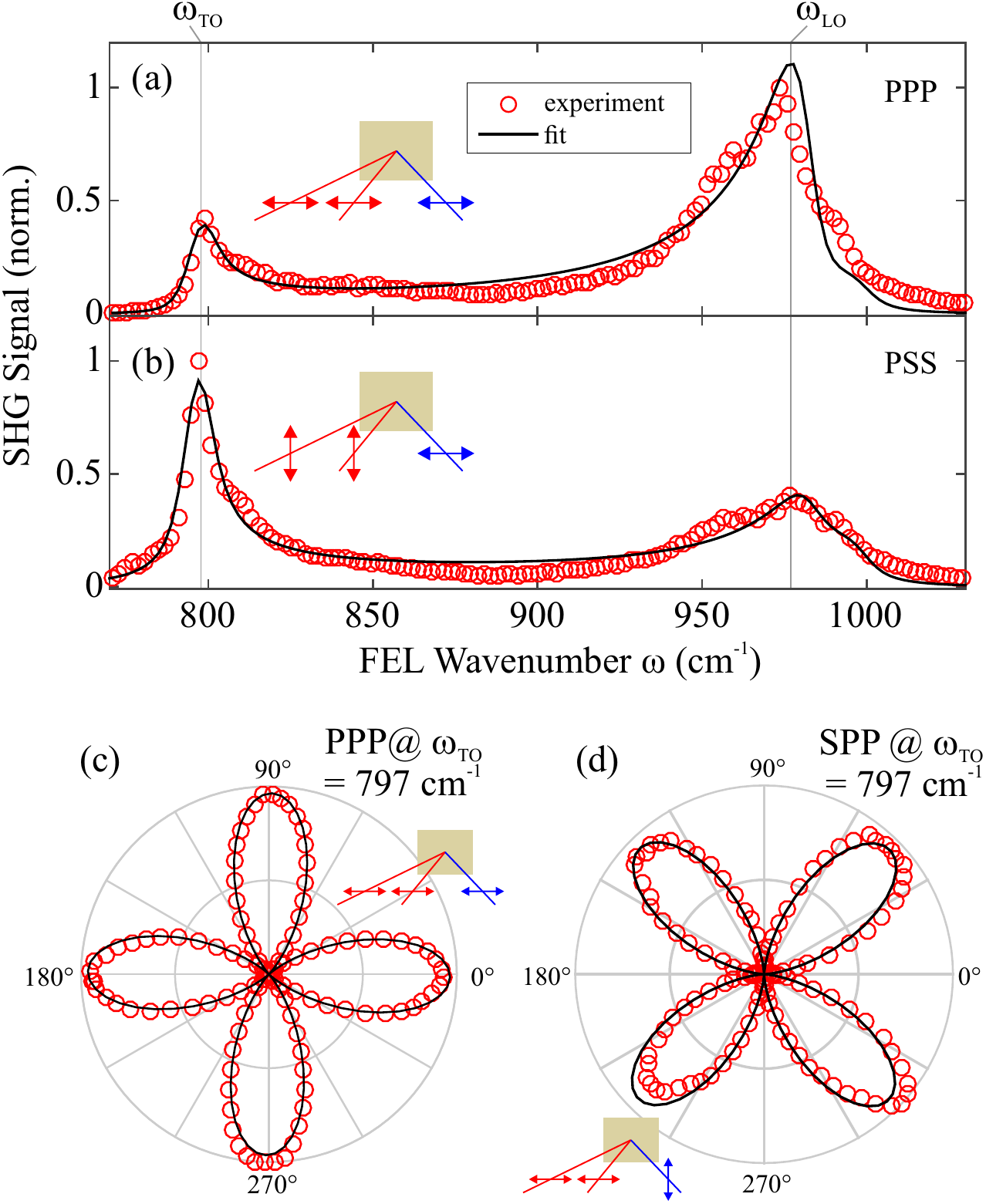}
\caption{Experimental SHG spectra of 3C-SiC for different polarization conditions: PPP (a) and PSS (b).  Two resonance features are observed in both spectra, coinciding with the fundamental optical phonon resonances, $\omega_{TO}$ and $\omega_{LO}$, respectively. The azimuthal behavior of the SHG response exhibits a four-fold symmetry as shown for the PPP (c) and SPP (d) geometries recorded at $\omega_{TO}$. The solid lines in all subplots are the result of a global fit of the $\chi^{(2)}$ line shape function (Eq.~\ref{eq:chi2}). The insets schematically depict the respective polarization geometry.}
\label{fig:exp.3C}
\end{figure}

In Fig.~\ref{fig:exp.3C} we show the experimental SHG spectra for polarization geometries PPP and PSS (a,b), as well as azimuthal scans for PPP and SPP geometries (c,d) for 3C-SiC. In the spectra, two SHG resonance features are observed at the fundamental zone-center optical phonon frequencies  $\omega_{TO}$ and $\omega_{LO}$, respectively, similar to previous results.\cite{Paarmann2015} The former at the TO phonon is due to the resonance in the nonlinear susceptibility $\chi^{(2)}$, while the latter originates from a resonance in the Fresnel transmission at the LO phonon frequency where $\epsilon$ is nearly zero.\cite{Paarmann2015} The solid lines in Fig.~\ref{fig:exp.3C} are the result of a global fit of the coefficients $C_{i,abc}$ in the $\chi^{(2)}$ line shape function (Eq.~\ref{eq:chi2}), that enters into the calculation of the SHG signals employing Eq.~\ref{eq:SHG.3C}. The experimental spectra also reproduce well the smooth SHG continuum between the two resonances as predicted by theory. 

The azimuthal SHG behavior for the FEL tuned to the TO phonon frequency $\omega_{TO} = 797$~cm$^{-1}$ is shown in Fig.~\ref{fig:exp.3C} (c) and (d), recorded in  PPP and SPP polarization geometries, respectively. The data clearly show the four-fold symmetry for both, as well as a phase shift of $\pi/4$ between the two geometries as expected for the zincblende crystal structure\cite{Yamada1994,Jordan1997} and predicted by Eq.~\ref{eq:SHG.3C}, where $I_{PPP}(2\omega)\propto|\sin(2\phi)|^2$ and $I_{SPP}(2\omega)\propto|\cos(2\phi)|^2$. Within signal-to-noise, we observe no isotropic contribution to these data, indicating that the signals are dominated by the bulk nonlinear response with no significant surface SHG signals.\cite{Yamada1994,Jordan1997} 


\subsection{Anisotropic 4H-SiC: c-cut ($\mathbf{c\parallel z}$)}

\begin{figure}[th]
\includegraphics[width=.6\textwidth]{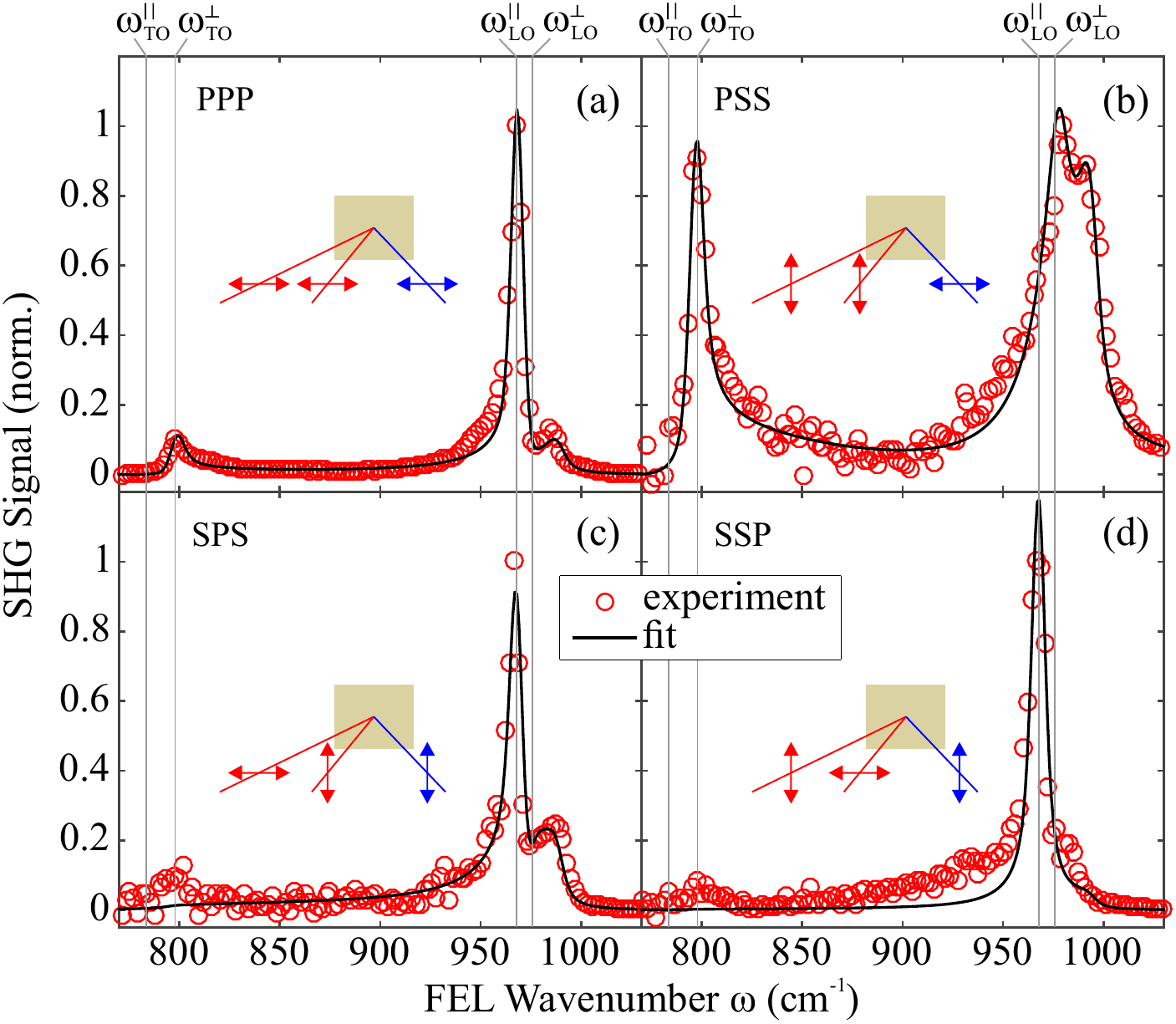}
\caption{Experimental SHG spectra of c-cut 4H-SiC for different polarization conditions: PPP (a), PSS (b), SPS (c), and SSP (d). All spectra involving P-polarized excitation exhibit a sharp peak at $\omega_{LO,\parallel}$. For SPS and SSP, the TO phonon resonance is suppressed since only linear resonance terms of the $\chi^{(2)}$ dispersion, Eq.~\ref{eq:chi2}, contribute to the signals. No azimuthal dependence is expected or observed for this sample.} 
\label{fig:exp.4Hc}
\end{figure}

We now turn the attention to the SHG response of c-cut 4H-SiC which is summarized in Fig.~\ref{fig:exp.4Hc}, where a frequency splitting between the axial and planar phonon resonances is expected. Due to the hexagonal crystal symmetry, no azimuthal variation of the SHG is predicted or observed. We recorded SHG spectra for all four polarization geometries with non-zero SHG signals: PPP (a), SPP (b), SPS (c), and SSP (d) which exhibit similar yet distinct SHG response. For all spectra, we also show the result of a fit of the $\chi^{(2)}$ line shape function entering into the SHG signal calculation exemplified by Eq.~\ref{eq:SHG.4Hc}. 

In all spectra, the largest SHG signals are observed in the LO phonon region of $960-980$~cm$^{-1}$. Specifically, all geometries involving at least one P-polarized excitation beam, PPP (a), SPS (c), and SSP (d), show a sharp and intense peak at $\omega_{LO,\parallel}$, i.e., the axial LO phonon resonance. This is a direct experimental proof for the  divergence of the anisotropy factor $\zeta = \epsilon_\perp/\epsilon_\parallel$ in the $z$-component of the Fresnel factor $L_{zz}$ in Eq.~\ref{eq:fresnel.4Hc}. The feature disappears for PSS geometry, where no $z$-components of the driving laser fields are involved. Here, we instead observe a spectrally shifted and softened double resonance, corresponding to spectral positions of the maxima of $L_{yy}$ for the two incidence angles.\cite{Paarmann2015} The weak high-energy shoulders clearly observable at $\omega \sim 980$~cm$^{-1}$ in Fig.~\ref{fig:exp.4Hc} (a,c,d) have the same origin.

In the TO phonon frequency region of $780-800$~cm$^{-1}$, resonant SHG enhancement is observed only for PPP and PSS geometries while it is largely suppressed for SPS and SSP. This can be understood by considering the anisotropy of the $\chi^{(2)}$ dispersion Eq.~\ref{eq:chi2}, i.e., the different resonance terms entering into each of the $\chi^{(2)}$ tensor elements, cf. Fig.~\ref{fig:sim} (b) for the $\chi^{(2)}$ dispersion extracted from a global fit to our measurements. Both PSS and PPP have a strong contribution from $\chi^{(2)}_{caa}$ which involves second order incoming planar TO resonances $1/D_\perp(\omega)$ in the second and third term of Eq.~\ref{eq:chi2}. These quadratic resonance terms are missing for $\chi^{(2)}_{aca}$ which is the sole tensor element contributing to SPS and SSP. Here, linear resonance terms $1/D_\perp(\omega)$ and $1/D_\parallel(\omega)$ in Eq.~\ref{eq:chi2} peak at different frequencies $\omega_{TO,\perp}$ and $\omega_{TO,\parallel}$, respectively. Additionally, the antiresonance of the Fresnel transmission of $L_{xx}$ at $\omega_{TO,\perp}$ shown in Fig.~\ref{fig:sim} (d) results in a suppression of the SHG output at $\omega_{TO,\perp}$. In fact, for SPS and SSP this Fresnel antiresonance exactly compensates the resonance in $\chi^{(2)}_{aca}$, resulting in a full suppression of the SHG enhancement at $\omega_{TO,\perp}$ for these geometries.  

None of the SHG spectra of c-cut 4H-SiC show SHG enhancement at the axial TO phonon at $\omega_{TO,\parallel} = 784$~cm$^{-1}$. This is particularly surprising for the PPP geometry, where the quadratic resonant enhancement of $\chi^{(2)}_{ccc}$ at the axial phonon frequency should contribute to the signal. However, here the SHG is again suppressed due to the pronounced antiresonance in the Fresnel factor $L_{zz}$, see Fig.~\ref{fig:sim} (d), also entering quadratically into the SHG response. Similar compensation effects suppress the SHG at  $\omega_{TO,\parallel}$ for SPS and SSP, while no enhancement of $\chi^{(2)}$ is expected for PSS at this frequency.


\subsection{Anisotropic 4H-SiC: a-cut ($\mathbf{c\perp z}$)}

\begin{figure*}[thb]
\includegraphics[width=\textwidth]{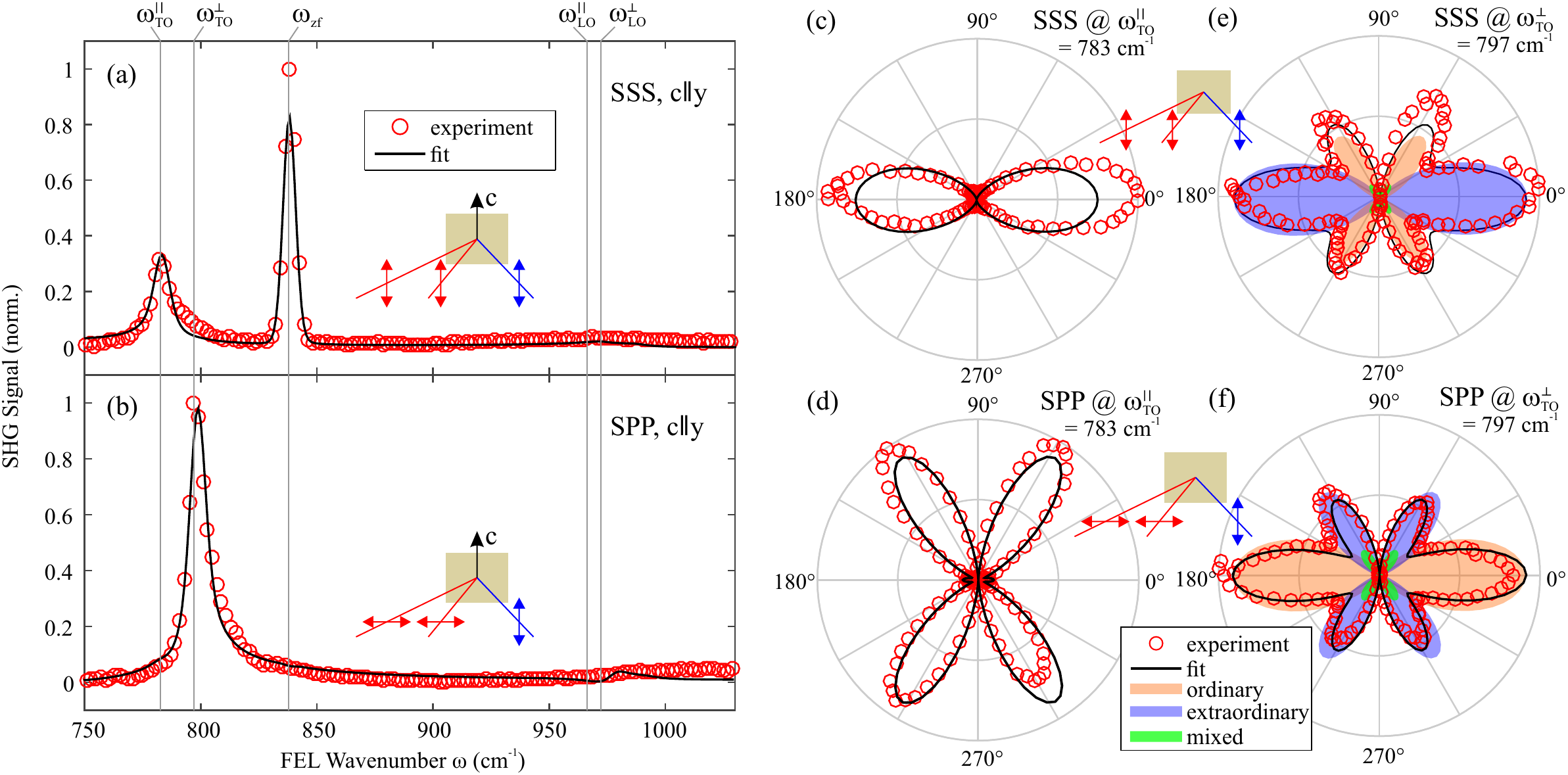}
\caption{Experimental SHG spectra (a,b) and azimuthal scans (c-f) of a-cut 4H-SiC. SHG Spectra are shown for the optical axis parallel to the $y$-axis, displaying purely extraordinary features for SSS polarization (a), and purely ordinary features for SPP (b). In particular, the weak mode at $\omega_{zf} = 838$~cm$^{-1}$ is observed as a narrow and intense SHG resonance in SSS geometry. Opposed to the previous samples, Fresnel transmission resonances in the LO frequency range are largely suppressed here. All azimuthal scans show a two-fold symmetry as the optical axis is rotated through the $x-y$ plane. Shown are representative scans taken at $\omega_{TO,\parallel}$ (c,d) and $\omega_{TO,\parallel}$ (e,f) for SSS (c,e) and SPP (d,f). Two distinct azimuthal shapes are observed (c,d), while additionally for (e,f) the orthogonal contributions from axial and planar phonon resonances overlap, as indicated by color-shaded areas representing the signal contributions from ordinary (orange), extraordinary (blue) and mixed (green) fundamental beams (e,f). In all plots, the solid lines show the result of a global fit of the $\chi^{(2)}$ line shape function, Eq.~\ref{eq:chi2}, used for calculating the SHG signals employing Eqs.~\ref{eq:SHG.4Ha} (a,b) and \ref{eq:SHG.4Ha.general} (c-f), respectively.}  
\label{fig:exp.4Ha}
\end{figure*}

This situation changes drastically for  a-cut 4H-SiC. Experimental SHG spectra for $c\parallel y$ and azimuthal scans at $\omega_{TO,\parallel}$ and $\omega_{TO,\perp}$  are shown in Fig.~\ref{fig:exp.4Ha}. Strikingly, the SHG spectra for SSS polarization geometry, Fig.~\ref{fig:exp.4Ha}~(a), now clearly exhibits a resonance when driving the axial TO phonon at $\omega_{TO,\parallel} = 784$~cm$^{-1}$. Additionally, an even more intense and narrow SHG resonance is observed at $\omega_{zf}=838$~cm$^{-1}$, i.e., when exciting the zone-folded weak mode of 4H-SiC. In fact, according to Eq.~\ref{eq:SHG.4Ha} only $\chi^{(2)}_{ccc}$ contributes to the SHG signal, making this measurement uniquely sensitive to axial resonances of the crystal. The large SHG signal at the weak mode frequency is surprising at first, since the oscillator strength of this mode is much lower as compared to the strong TO modes, i.e., one would also expect a weaker $\chi^{(2)}$ resonance. However, the Fresnel suppression acting on the strong phonon resonances is essentially absent for the zone-folded mode. This is because the amplitude of the $\epsilon_\parallel$-resonance at $\omega_{zf}$ is much smaller than the large negative permittivity offset in the middle of the Reststrahl. In consequence, the smaller yet full amplitude of the $\chi^{(2)}$-resonance is carried over to the SHG signals for this weak mode. 

The SHG spectrum for SPP geometry, on the other hand, is sensitive only to planar phonons, as evidenced by a single SHG resonance at $\omega_{TO,\perp}=797$~cm$^{-1}$.  By inspection of Eq.~\ref{eq:SHG.4Ha}, only $\chi^{(2)}_{caa}$ contributes to the SHG signal for this geometry, i.e., all incoming resonance terms of $\chi^{(2)}_{caa}$ in Eq.~\ref{eq:chi2} occur at $\omega_{TO,\perp}$. None of the SHG spectra of the a-cut 4H-SiC sample show any resonances in the LO frequency region, which we attribute to graphite contamination of the sample surface.

The azimuthal shapes exhibit a two-fold symmetry as the optical axis is rotated about the surface normal. We observe two unique and complementary azimuthal patterns exemplified in Fig.~\ref{fig:exp.4Ha} (c) and (d) for SSS and SPP geometries, respectively, where the FEL is tuned to $\omega_{TO,\parallel}=784$~cm$^{-1}$. For SSS, we observe two peaks, each for $c\parallel y$, while the signal is suppressed for $c\parallel x$. For SPP, signals are suppressed for all main axes but recover for intermediate angles $|\angle(c,y)|\sim35^\circ$. The azimuthal shapes are identical when driving the zone-folded weak mode at $\omega_{zf} = 838$~cm$^{-1}$. Also, for both axial phonon resonances, the identical patterns are observed for PPP and PSS, but rotated by $90^\circ$ (not shown).

In contrast, when tuning the FEL to the planar TO resonance at $\omega_{TO,\perp} = 797$~cm$^{-1}$ we observe very similar azimuthal patterns for the same SSS and SPP geometries, see Fig.~\ref{fig:exp.4Ha} (e) and (f). However, this can be understood quite easily by interpreting these data as a superposition of the two unique patterns, cf. Fig.~\ref{fig:exp.4Ha} (c,d), one originating from the tail of the axial (extraordinary) phonon resonance and the other from the planar (ordinary) phonon resonance. To illustrate the effect, we show the calculated ordinary, extraordinary and mixed beam (one beam ordinary, one beam extraordinary) contributions to the total azimuthal pattern in Fig.~\ref{fig:exp.4Ha} (e,f), where clearly the overlapping contributions can be traced back to the respective phonon resonances. This effect occurs for both SSS and SPP geometries, just with opposite amplitude ratios of the two contributions, surprisingly leading to almost identical shapes. Again, we observe identical azimuthal patterns for PPP and PSS, but rotated by 90$^\circ$ as compared to SSS and SPP, respectively (not shown). This high symmetry of the experimental data puts very specific constrains on the relative magnitudes of the contributing $\chi^{(2)}$ tensor elements, thus allowing for a global fit of the $\chi^{(2)}$ line shapes shown in Fig.~\ref{fig:sim}.

\section{Discussion and Conclusion}

\begin{figure*}[bht]
\includegraphics[width=.7\textwidth]{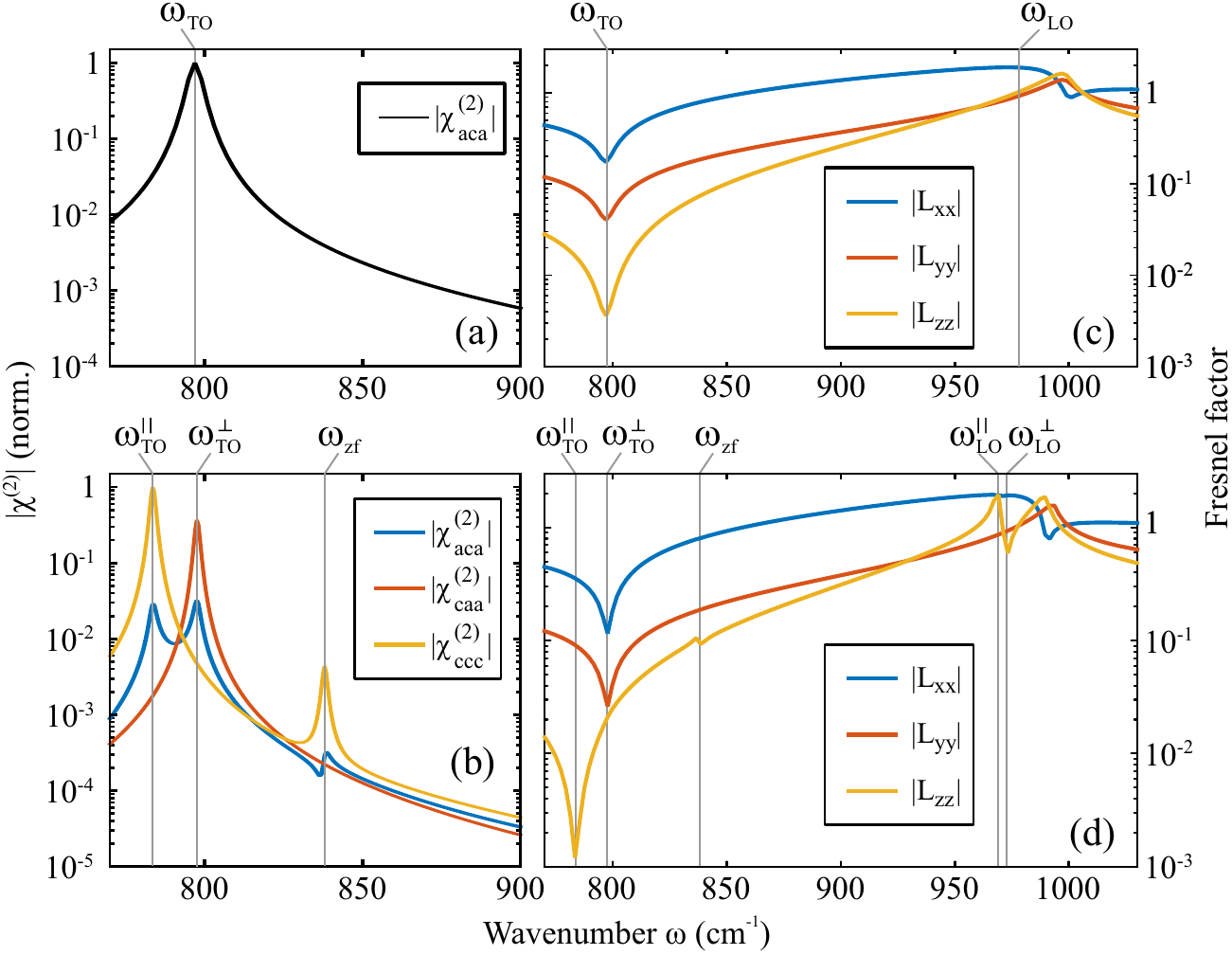}
\caption{Dispersion of the nonlinear susceptibility $\chi^{(2)}$ extracted from global fits of the experimental SHG data for 3C-SiC (a) and a-cut 4H-SiC (b). Distinct anisotropic features in (b) appear as peak shifts and splittings for the different tensor components, following the respective TO phonon frequencies $\omega_{TO,\parallel}$ and $\omega_{TO,\perp}$, as well as an additional resonance at the zone-folded mode frequency. For comparison, we also show the calculated Fresnel transmission coefficients for 3C-SiC (c) and c-cut 4H-SiC (d) for an incidence angle of $\theta^{i} = 60^\circ$. In the latter, the crystal anistropy results in a shift of the anti-resonance at the TO frequency and an additional sharp feature at the LO frequencies for $L_{zz}$ (yellow line). Please note that all plots are shown on a logarithmic scale.} 
\label{fig:sim}
\end{figure*}

In the previous sections, we have identified multiple pronounced effects of the crystal anisotropy on the SHG response of SiC which are summarized in Fig.~\ref{fig:sim}. Here, we compare the dispersion of the two important ingredients of the nonlinear-optical response in Eq.~\ref{eq:PSHG}, i.e., the nonlinear susceptibility $\chi^{(2)}$ and the local field factors represented by the Fresnel transmission $L$, for cubic and hexagonal SiC. For $\chi^{(2)}$, the change of the point group symmetry, $\bar 43m$ to $6mm$,  increases the number of unique non-zero elements of the tensor from one to three. At the same time, the anisotropy in the phonon frequencies, planar vs. axial, results in significantly different dispersions of these three tensor elements. For incoming resonances at $\omega$, $\chi^{(2)}_{caa}$ and $\chi^{(2)}_{ccc}$ uniquely carry information on the planar and axial phonon resonances, respectively, while $\chi^{(2)}_{aca}$ is essentially a coupling term that only contributes to the SHG signal if both unique crystal axes provide a resonant response. We here do not investigate outgoing resonances,\cite{Dekorsy2003}i.e., for $\omega \approx \omega_{TO}/2$, where also $\chi^{(2)}_{caa}$ is expected to be sensitive to axial phonon resonances.

A surprisingly strong anisotropic effect is observed in the Fresnel factors, cf. Fig.~\ref{fig:sim} (c) and (d), where we compare the calculations for 3C-SiC and c-cut 4H-SiC, respectively. While the in-plane local fields described by $L_{xx}$ and $L_{yy}$ are essentially identical for both crystals, a pronounced anisotropic feature appears in $L_{zz}$ at $\omega_{LO,\parallel}$ for 4H-SiC. The anisotropy factor $\zeta = \epsilon_{\perp}/\epsilon_{\parallel}$ in Eq.~\ref{eq:fresnel.4Hc} essentially results in a divergence and prominent enhancement of the fields inside the crystal when excited at this frequency. The experimental SHG spectra are very sensitive to this effect, exhibiting a sharp and intense peak at this resonance whenever $L_{zz}$ is involved in the signal generation.

In the work presented here, the use of different crystal orientations, i.e., c-cut and a-cut 4H-SiC, allowed to clearly bring out the different anisotropy effects which are otherwise hidden due to counteracting $\chi^{(2)}$ enhancement and Fresnel suppression mechanisms. Most importantly, it was possible to find a suitable combination of sample and polarization geometry to clearly separate the anisotropic features in the nonlinear response, while the azimuthal behavior at different resonance positions facilitates identification as well as quantitative comparison of these contributions. As a consequence, we were able to observe a strong SHG resonance feature at the zone-folded weak mode of 4H-SiC, cf. Fig.~\ref{fig:exp.4Ha} (a), which in fact dominates the SHG spectra in these cases. We note that the latter result is quite promising since it allows to observe strong nonlinear signals from weak or dilute oscillators, as long as they are spectrally contained in the Reststrahl. This is owing to the large negative permittivity in the Reststrahl, essentially disabling  the Fresnel suppression acting on the strong phonon resonances, and thereby enhancing the SHG signal particularly for weak oscillators.

Overall, we find that mid-IR SHG spectroscopy is an excellent tool to investigate and identify even subtle effects due to the crystal anisotropy with high sensitivity. The special cases we outlined can be generalized to successfully implement nonlinear-optical experiments in other Reststrahl materials, which often exhibit pronounced effects due to crystal anisotropy. In fact, the procedure presented here is directly applicable to other materials with the same $6mm$ point group symmetry, as for instance all crystals occurring in the wurtzite structure such as AlN, ZnO, CdSe, and ZnTe, to name a few. All these crystals are also polar dielectrics with a Reststrahl region, and are often very efficient nonlinear materials. Similar recipes are expected to be applicable to uniaxial crystals of different symmetries such as $\alpha$-SiO$_2$. In more general terms, it may be possible to specifically design artificial materials in order to actively enhance the nonlinear-optical response at specific anisotropic resonances. In fact, recent experiments show resonant interactions between sub-diffractional, localized surface phonon polaritons and zone-folded weak modes in SiC nanopillars resulting in a strong enhancement of the SHG response.\cite{Razdolski2016}

In summary, we studied the effects of crystal anisotropy on the SHG response of SiC in its mid-IR Reststrahl spectral region. We find distinct features in the signals that, with the help of the theoretical model developed here and by comparison of the experimental results of isotropic and anisotropic SiC samples of different crystal cuts, can be assigned to originate only from the crystal anisotropy. We experimentally verify the planar and axial phonon resonances appearing selectively in the dispersion of the corresponding tensor components of the nonlinear susceptibility  $\chi^{(2)}$, including a pronounced resonance at the zone-folded weak phonon frequency in $\chi^{(2)}_{ccc}$. Similarly, we identify an anisotropy factor $\zeta = \epsilon_\perp/\epsilon_\parallel$ in the Fresnel transmission which leads to a steep enhancement of the local field amplitudes spectrally locked to the axial LO frequency. The same behavior is expected for the large material class of wurtzite-type dielectrics. The anisotropic resonances can be utilized to selectively enhance the nonlinear-optical response, for instance using surface phonon polaritons in anisotropic polar dielectrics.

The authors thank J.D. Caldwell (NRL, Washington D.C.), R. Koch (TU Chemnitz), and K. Horn (FHI Berlin) for providing the SiC samples.


\begin{thebibliography}{34}%
\makeatletter
\providecommand \@ifxundefined [1]{%
 \@ifx{#1\undefined}
}%
\providecommand \@ifnum [1]{%
 \ifnum #1\expandafter \@firstoftwo
 \else \expandafter \@secondoftwo
 \fi
}%
\providecommand \@ifx [1]{%
 \ifx #1\expandafter \@firstoftwo
 \else \expandafter \@secondoftwo
 \fi
}%
\providecommand \natexlab [1]{#1}%
\providecommand \enquote  [1]{``#1''}%
\providecommand \bibnamefont  [1]{#1}%
\providecommand \bibfnamefont [1]{#1}%
\providecommand \citenamefont [1]{#1}%
\providecommand \href@noop [0]{\@secondoftwo}%
\providecommand \href [0]{\begingroup \@sanitize@url \@href}%
\providecommand \@href[1]{\@@startlink{#1}\@@href}%
\providecommand \@@href[1]{\endgroup#1\@@endlink}%
\providecommand \@sanitize@url [0]{\catcode `\\12\catcode `\$12\catcode
  `\&12\catcode `\#12\catcode `\^12\catcode `\_12\catcode `\%12\relax}%
\providecommand \@@startlink[1]{}%
\providecommand \@@endlink[0]{}%
\providecommand \url  [0]{\begingroup\@sanitize@url \@url }%
\providecommand \@url [1]{\endgroup\@href {#1}{\urlprefix }}%
\providecommand \urlprefix  [0]{URL }%
\providecommand \Eprint [0]{\href }%
\providecommand \doibase [0]{http://dx.doi.org/}%
\providecommand \selectlanguage [0]{\@gobble}%
\providecommand \bibinfo  [0]{\@secondoftwo}%
\providecommand \bibfield  [0]{\@secondoftwo}%
\providecommand \translation [1]{[#1]}%
\providecommand \BibitemOpen [0]{}%
\providecommand \bibitemStop [0]{}%
\providecommand \bibitemNoStop [0]{.\EOS\space}%
\providecommand \EOS [0]{\spacefactor3000\relax}%
\providecommand \BibitemShut  [1]{\csname bibitem#1\endcsname}%
\let\auto@bib@innerbib\@empty
\bibitem [{\citenamefont {Taubner}\ \emph {et~al.}(2006)\citenamefont
  {Taubner}, \citenamefont {Korobkin}, \citenamefont {Urzhumov}, \citenamefont
  {Shvets},\ and\ \citenamefont {Hillenbrand}}]{Taubner2006}%
  \BibitemOpen
  \bibfield  {author} {\bibinfo {author} {\bibfnamefont {T.}~\bibnamefont
  {Taubner}}, \bibinfo {author} {\bibfnamefont {D.}~\bibnamefont {Korobkin}},
  \bibinfo {author} {\bibfnamefont {Y.}~\bibnamefont {Urzhumov}}, \bibinfo
  {author} {\bibfnamefont {G.}~\bibnamefont {Shvets}}, \ and\ \bibinfo {author}
  {\bibfnamefont {R.}~\bibnamefont {Hillenbrand}},\ }\href {\doibase
  10.1126/science.1131025} {\bibfield  {journal} {\bibinfo  {journal} {Science
  (New York, N.Y.)}\ }\textbf {\bibinfo {volume} {313}},\ \bibinfo {pages}
  {1595} (\bibinfo {year} {2006})}\BibitemShut {NoStop}%
\bibitem [{\citenamefont {Caldwell}\ \emph {et~al.}(2013)\citenamefont
  {Caldwell}, \citenamefont {Glembocki}, \citenamefont {Francescato},
  \citenamefont {Sharac}, \citenamefont {Giannini}, \citenamefont {Bezares},
  \citenamefont {Long}, \citenamefont {Owrutsky}, \citenamefont {Vurgaftman},
  \citenamefont {Tischler}, \citenamefont {Wheeler}, \citenamefont {Bassim},
  \citenamefont {Shirey}, \citenamefont {Kasica},\ and\ \citenamefont
  {Maier}}]{Caldwell2013}%
  \BibitemOpen
  \bibfield  {author} {\bibinfo {author} {\bibfnamefont {J.~D.}\ \bibnamefont
  {Caldwell}}, \bibinfo {author} {\bibfnamefont {O.~J.}\ \bibnamefont
  {Glembocki}}, \bibinfo {author} {\bibfnamefont {Y.}~\bibnamefont
  {Francescato}}, \bibinfo {author} {\bibfnamefont {N.}~\bibnamefont {Sharac}},
  \bibinfo {author} {\bibfnamefont {V.}~\bibnamefont {Giannini}}, \bibinfo
  {author} {\bibfnamefont {F.~J.}\ \bibnamefont {Bezares}}, \bibinfo {author}
  {\bibfnamefont {J.~P.}\ \bibnamefont {Long}}, \bibinfo {author}
  {\bibfnamefont {J.~C.}\ \bibnamefont {Owrutsky}}, \bibinfo {author}
  {\bibfnamefont {I.}~\bibnamefont {Vurgaftman}}, \bibinfo {author}
  {\bibfnamefont {J.~G.}\ \bibnamefont {Tischler}}, \bibinfo {author}
  {\bibfnamefont {V.~D.}\ \bibnamefont {Wheeler}}, \bibinfo {author}
  {\bibfnamefont {N.~D.}\ \bibnamefont {Bassim}}, \bibinfo {author}
  {\bibfnamefont {L.~M.}\ \bibnamefont {Shirey}}, \bibinfo {author}
  {\bibfnamefont {R.}~\bibnamefont {Kasica}}, \ and\ \bibinfo {author}
  {\bibfnamefont {S.~A.}\ \bibnamefont {Maier}},\ }\href {\doibase
  10.1021/nl401590g} {\bibfield  {journal} {\bibinfo  {journal} {Nano Letters}\
  }\textbf {\bibinfo {volume} {13}},\ \bibinfo {pages} {3690} (\bibinfo {year}
  {2013})}\BibitemShut {NoStop}%
\bibitem [{\citenamefont {Gubbin}\ \emph {et~al.}(2016)\citenamefont {Gubbin},
  \citenamefont {Martini}, \citenamefont {Politi}, \citenamefont {Maier},\ and\
  \citenamefont {{De Liberato}}}]{Gubbin2016}%
  \BibitemOpen
  \bibfield  {author} {\bibinfo {author} {\bibfnamefont {C.~R.}\ \bibnamefont
  {Gubbin}}, \bibinfo {author} {\bibfnamefont {F.}~\bibnamefont {Martini}},
  \bibinfo {author} {\bibfnamefont {A.}~\bibnamefont {Politi}}, \bibinfo
  {author} {\bibfnamefont {S.~A.}\ \bibnamefont {Maier}}, \ and\ \bibinfo
  {author} {\bibfnamefont {S.}~\bibnamefont {{De Liberato}}},\ }\href {\doibase
  10.1103/PhysRevLett.116.246402} {\bibfield  {journal} {\bibinfo  {journal}
  {Physical Review Letters}\ }\textbf {\bibinfo {volume} {116}},\ \bibinfo
  {pages} {246402} (\bibinfo {year} {2016})}\BibitemShut {NoStop}%
\bibitem [{\citenamefont {Li}\ \emph {et~al.}(2016)\citenamefont {Li},
  \citenamefont {Yang}, \citenamefont {Ma{\ss}}, \citenamefont {Hanss},
  \citenamefont {Lewin}, \citenamefont {Michel}, \citenamefont {Wuttig},\ and\
  \citenamefont {Taubner}}]{Li2016}%
  \BibitemOpen
  \bibfield  {author} {\bibinfo {author} {\bibfnamefont {P.}~\bibnamefont
  {Li}}, \bibinfo {author} {\bibfnamefont {X.}~\bibnamefont {Yang}}, \bibinfo
  {author} {\bibfnamefont {T.~W.~W.}\ \bibnamefont {Ma{\ss}}}, \bibinfo
  {author} {\bibfnamefont {J.}~\bibnamefont {Hanss}}, \bibinfo {author}
  {\bibfnamefont {M.}~\bibnamefont {Lewin}}, \bibinfo {author} {\bibfnamefont
  {A.-K.~U.}\ \bibnamefont {Michel}}, \bibinfo {author} {\bibfnamefont
  {M.}~\bibnamefont {Wuttig}}, \ and\ \bibinfo {author} {\bibfnamefont
  {T.}~\bibnamefont {Taubner}},\ }\href {\doibase 10.1038/nmat4649} {\bibfield
  {journal} {\bibinfo  {journal} {Nature Materials}\ } (\bibinfo {year}
  {2016}),\ 10.1038/nmat4649}\BibitemShut {NoStop}%
\bibitem [{\citenamefont {Caldwell}\ \emph {et~al.}(2015)\citenamefont
  {Caldwell}, \citenamefont {Lindsay}, \citenamefont {Giannini}, \citenamefont
  {Vurgaftman}, \citenamefont {Reinecke}, \citenamefont {Maier},\ and\
  \citenamefont {Glembocki}}]{Caldwell2014a}%
  \BibitemOpen
  \bibfield  {author} {\bibinfo {author} {\bibfnamefont {J.~D.}\ \bibnamefont
  {Caldwell}}, \bibinfo {author} {\bibfnamefont {L.}~\bibnamefont {Lindsay}},
  \bibinfo {author} {\bibfnamefont {V.}~\bibnamefont {Giannini}}, \bibinfo
  {author} {\bibfnamefont {I.}~\bibnamefont {Vurgaftman}}, \bibinfo {author}
  {\bibfnamefont {T.~L.}\ \bibnamefont {Reinecke}}, \bibinfo {author}
  {\bibfnamefont {S.~A.}\ \bibnamefont {Maier}}, \ and\ \bibinfo {author}
  {\bibfnamefont {O.~J.}\ \bibnamefont {Glembocki}},\ }\href {\doibase
  10.1515/nanoph-2014-0003} {\bibfield  {journal} {\bibinfo  {journal}
  {Nanophotonics}\ }\textbf {\bibinfo {volume} {4}},\ \bibinfo {pages} {1}
  (\bibinfo {year} {2015})}\BibitemShut {NoStop}%
\bibitem [{\citenamefont {Dai}\ \emph {et~al.}(2014)\citenamefont {Dai},
  \citenamefont {Fei}, \citenamefont {Ma}, \citenamefont {Rodin}, \citenamefont
  {Wagner}, \citenamefont {McLeod}, \citenamefont {Liu}, \citenamefont
  {Gannett}, \citenamefont {Regan}, \citenamefont {Watanabe}, \citenamefont
  {Taniguchi}, \citenamefont {Thiemens}, \citenamefont {Dominguez},
  \citenamefont {Neto}, \citenamefont {Zettl}, \citenamefont {Keilmann},
  \citenamefont {Jarillo-Herrero}, \citenamefont {Fogler},\ and\ \citenamefont
  {Basov}}]{Dai2014}%
  \BibitemOpen
  \bibfield  {author} {\bibinfo {author} {\bibfnamefont {S.}~\bibnamefont
  {Dai}}, \bibinfo {author} {\bibfnamefont {Z.}~\bibnamefont {Fei}}, \bibinfo
  {author} {\bibfnamefont {Q.}~\bibnamefont {Ma}}, \bibinfo {author}
  {\bibfnamefont {A.~S.}\ \bibnamefont {Rodin}}, \bibinfo {author}
  {\bibfnamefont {M.}~\bibnamefont {Wagner}}, \bibinfo {author} {\bibfnamefont
  {A.~S.}\ \bibnamefont {McLeod}}, \bibinfo {author} {\bibfnamefont {M.~K.}\
  \bibnamefont {Liu}}, \bibinfo {author} {\bibfnamefont {W.}~\bibnamefont
  {Gannett}}, \bibinfo {author} {\bibfnamefont {W.}~\bibnamefont {Regan}},
  \bibinfo {author} {\bibfnamefont {K.}~\bibnamefont {Watanabe}}, \bibinfo
  {author} {\bibfnamefont {T.}~\bibnamefont {Taniguchi}}, \bibinfo {author}
  {\bibfnamefont {M.}~\bibnamefont {Thiemens}}, \bibinfo {author}
  {\bibfnamefont {G.}~\bibnamefont {Dominguez}}, \bibinfo {author}
  {\bibfnamefont {A.~H.~C.}\ \bibnamefont {Neto}}, \bibinfo {author}
  {\bibfnamefont {A.}~\bibnamefont {Zettl}}, \bibinfo {author} {\bibfnamefont
  {F.}~\bibnamefont {Keilmann}}, \bibinfo {author} {\bibfnamefont
  {P.}~\bibnamefont {Jarillo-Herrero}}, \bibinfo {author} {\bibfnamefont
  {M.~M.}\ \bibnamefont {Fogler}}, \ and\ \bibinfo {author} {\bibfnamefont
  {D.~N.}\ \bibnamefont {Basov}},\ }\href {\doibase 10.1126/science.1246833}
  {\bibfield  {journal} {\bibinfo  {journal} {Science}\ }\textbf {\bibinfo
  {volume} {343}},\ \bibinfo {pages} {1125} (\bibinfo {year}
  {2014})}\BibitemShut {NoStop}%
\bibitem [{\citenamefont {Yoxall}\ \emph {et~al.}(2015)\citenamefont {Yoxall},
  \citenamefont {Schnell}, \citenamefont {Nikitin}, \citenamefont {Txoperena},
  \citenamefont {Woessner}, \citenamefont {Lundeberg}, \citenamefont
  {Casanova}, \citenamefont {Hueso}, \citenamefont {Koppens},\ and\
  \citenamefont {Hillenbrand}}]{Yoxall2015}%
  \BibitemOpen
  \bibfield  {author} {\bibinfo {author} {\bibfnamefont {E.}~\bibnamefont
  {Yoxall}}, \bibinfo {author} {\bibfnamefont {M.}~\bibnamefont {Schnell}},
  \bibinfo {author} {\bibfnamefont {A.~Y.}\ \bibnamefont {Nikitin}}, \bibinfo
  {author} {\bibfnamefont {O.}~\bibnamefont {Txoperena}}, \bibinfo {author}
  {\bibfnamefont {A.}~\bibnamefont {Woessner}}, \bibinfo {author}
  {\bibfnamefont {M.~B.}\ \bibnamefont {Lundeberg}}, \bibinfo {author}
  {\bibfnamefont {F.}~\bibnamefont {Casanova}}, \bibinfo {author}
  {\bibfnamefont {L.~E.}\ \bibnamefont {Hueso}}, \bibinfo {author}
  {\bibfnamefont {F.~H.~L.}\ \bibnamefont {Koppens}}, \ and\ \bibinfo {author}
  {\bibfnamefont {R.}~\bibnamefont {Hillenbrand}},\ }\href {\doibase
  10.1038/nphoton.2015.166} {\bibfield  {journal} {\bibinfo  {journal} {Nature
  Photonics}\ }\textbf {\bibinfo {volume} {9}},\ \bibinfo {pages} {674}
  (\bibinfo {year} {2015})}\BibitemShut {NoStop}%
\bibitem [{\citenamefont {Caldwell}\ \emph {et~al.}(2014)\citenamefont
  {Caldwell}, \citenamefont {Kretinin}, \citenamefont {Chen}, \citenamefont
  {Giannini}, \citenamefont {Fogler}, \citenamefont {Francescato},
  \citenamefont {Ellis}, \citenamefont {Tischler}, \citenamefont {Woods},
  \citenamefont {Giles}, \citenamefont {Hong}, \citenamefont {Watanabe},
  \citenamefont {Taniguchi}, \citenamefont {Maier},\ and\ \citenamefont
  {Novoselov}}]{Caldwell2014}%
  \BibitemOpen
  \bibfield  {author} {\bibinfo {author} {\bibfnamefont {J.~D.}\ \bibnamefont
  {Caldwell}}, \bibinfo {author} {\bibfnamefont {A.~V.}\ \bibnamefont
  {Kretinin}}, \bibinfo {author} {\bibfnamefont {Y.}~\bibnamefont {Chen}},
  \bibinfo {author} {\bibfnamefont {V.}~\bibnamefont {Giannini}}, \bibinfo
  {author} {\bibfnamefont {M.~M.}\ \bibnamefont {Fogler}}, \bibinfo {author}
  {\bibfnamefont {Y.}~\bibnamefont {Francescato}}, \bibinfo {author}
  {\bibfnamefont {C.~T.}\ \bibnamefont {Ellis}}, \bibinfo {author}
  {\bibfnamefont {J.~G.}\ \bibnamefont {Tischler}}, \bibinfo {author}
  {\bibfnamefont {C.~R.}\ \bibnamefont {Woods}}, \bibinfo {author}
  {\bibfnamefont {A.~J.}\ \bibnamefont {Giles}}, \bibinfo {author}
  {\bibfnamefont {M.}~\bibnamefont {Hong}}, \bibinfo {author} {\bibfnamefont
  {K.}~\bibnamefont {Watanabe}}, \bibinfo {author} {\bibfnamefont
  {T.}~\bibnamefont {Taniguchi}}, \bibinfo {author} {\bibfnamefont {S.~a.}\
  \bibnamefont {Maier}}, \ and\ \bibinfo {author} {\bibfnamefont {K.~S.}\
  \bibnamefont {Novoselov}},\ }\href {\doibase 10.1038/ncomms6221} {\bibfield
  {journal} {\bibinfo  {journal} {Nature communications}\ }\textbf {\bibinfo
  {volume} {5}},\ \bibinfo {pages} {5221} (\bibinfo {year} {2014})}\BibitemShut
  {NoStop}%
\bibitem [{\citenamefont {Nakashima}\ and\ \citenamefont
  {Harima}(1997)}]{Nakashima1997}%
  \BibitemOpen
  \bibfield  {author} {\bibinfo {author} {\bibfnamefont {S.}~\bibnamefont
  {Nakashima}}\ and\ \bibinfo {author} {\bibfnamefont {H.}~\bibnamefont
  {Harima}},\ }\href
  {http://onlinelibrary.wiley.com/doi/10.1002/1521-396X(199707)162:1{\%}3C39::AID-PSSA39{\%}3E3.0.CO;2-L/abstract}
  {\bibfield  {journal} {\bibinfo  {journal} {physica status solidi (a)}\
  }\textbf {\bibinfo {volume} {162}},\ \bibinfo {pages} {39} (\bibinfo {year}
  {1997})}\BibitemShut {NoStop}%
\bibitem [{\citenamefont {Mutschke}\ \emph {et~al.}(1999)\citenamefont
  {Mutschke}, \citenamefont {Andersen}, \citenamefont {Clement}, \citenamefont
  {Henning},\ and\ \citenamefont {Peiter}}]{Mutschke1999}%
  \BibitemOpen
  \bibfield  {author} {\bibinfo {author} {\bibfnamefont {H.}~\bibnamefont
  {Mutschke}}, \bibinfo {author} {\bibfnamefont {A.~C.~A.}\ \bibnamefont
  {Andersen}}, \bibinfo {author} {\bibfnamefont {D.}~\bibnamefont {Clement}},
  \bibinfo {author} {\bibfnamefont {T.}~\bibnamefont {Henning}}, \ and\
  \bibinfo {author} {\bibfnamefont {G.}~\bibnamefont {Peiter}},\ }\href
  {http://aa.springer.de/papers/9345001/2300187/small.htm} {\bibfield
  {journal} {\bibinfo  {journal} {Astron. Astrophys.}\ }\textbf {\bibinfo
  {volume} {345}},\ \bibinfo {pages} {187} (\bibinfo {year}
  {1999})}\BibitemShut {NoStop}%
\bibitem [{\citenamefont {Engelbrecht}\ and\ \citenamefont
  {Helbig}(1993)}]{Engelbrecht1993}%
  \BibitemOpen
  \bibfield  {author} {\bibinfo {author} {\bibfnamefont {F.}~\bibnamefont
  {Engelbrecht}}\ and\ \bibinfo {author} {\bibfnamefont {R.}~\bibnamefont
  {Helbig}},\ }\href {\doibase 10.1103/PhysRevB.48.15698} {\bibfield  {journal}
  {\bibinfo  {journal} {Physical Review B}\ }\textbf {\bibinfo {volume} {48}},\
  \bibinfo {pages} {15698} (\bibinfo {year} {1993})}\BibitemShut {NoStop}%
\bibitem [{\citenamefont {Bluet}\ \emph {et~al.}(1999)\citenamefont {Bluet},
  \citenamefont {Chourou}, \citenamefont {Anikin},\ and\ \citenamefont
  {Madar}}]{Bluet1999}%
  \BibitemOpen
  \bibfield  {author} {\bibinfo {author} {\bibfnamefont {J.}~\bibnamefont
  {Bluet}}, \bibinfo {author} {\bibfnamefont {K.}~\bibnamefont {Chourou}},
  \bibinfo {author} {\bibfnamefont {M.}~\bibnamefont {Anikin}}, \ and\ \bibinfo
  {author} {\bibfnamefont {R.}~\bibnamefont {Madar}},\ }\href {\doibase
  10.1016/S0921-5107(98)00504-2} {\bibfield  {journal} {\bibinfo  {journal}
  {Materials Science and Engineering: B}\ }\textbf {\bibinfo {volume}
  {61-62}},\ \bibinfo {pages} {212} (\bibinfo {year} {1999})}\BibitemShut
  {NoStop}%
\bibitem [{\citenamefont {Caldwell}\ \emph {et~al.}(2016)\citenamefont
  {Caldwell}, \citenamefont {Vurgaftman}, \citenamefont {Tischler},
  \citenamefont {Glembocki}, \citenamefont {Owrutsky},\ and\ \citenamefont
  {Reinecke}}]{Caldwell2016}%
  \BibitemOpen
  \bibfield  {author} {\bibinfo {author} {\bibfnamefont {J.~D.}\ \bibnamefont
  {Caldwell}}, \bibinfo {author} {\bibfnamefont {I.}~\bibnamefont
  {Vurgaftman}}, \bibinfo {author} {\bibfnamefont {J.~G.}\ \bibnamefont
  {Tischler}}, \bibinfo {author} {\bibfnamefont {O.~J.}\ \bibnamefont
  {Glembocki}}, \bibinfo {author} {\bibfnamefont {J.~C.}\ \bibnamefont
  {Owrutsky}}, \ and\ \bibinfo {author} {\bibfnamefont {T.~L.}\ \bibnamefont
  {Reinecke}},\ }\href {\doibase 10.1038/nnano.2015.305} {\bibfield  {journal}
  {\bibinfo  {journal} {Nature Nanotechnology}\ }\textbf {\bibinfo {volume}
  {11}},\ \bibinfo {pages} {9} (\bibinfo {year} {2016})}\BibitemShut {NoStop}%
\bibitem [{\citenamefont {Burfoot}\ and\ \citenamefont
  {Tayler}(1979)}]{Burfoot1979}%
  \BibitemOpen
  \bibfield  {author} {\bibinfo {author} {\bibfnamefont {J.~C.}\ \bibnamefont
  {Burfoot}}\ and\ \bibinfo {author} {\bibfnamefont {G.~W.}\ \bibnamefont
  {Tayler}},\ }\href@noop {} {\emph {\bibinfo {title} {{Polar dielectrics and
  their applications}}}}\ (\bibinfo  {publisher} {University of Califonia
  Press},\ \bibinfo {year} {1979})\ p.\ \bibinfo {pages} {465}\BibitemShut
  {NoStop}%
\bibitem [{\citenamefont {Shen}(1994)}]{Shen1994}%
  \BibitemOpen
  \bibfield  {author} {\bibinfo {author} {\bibfnamefont {Y.~R.}\ \bibnamefont
  {Shen}},\ }\href {\doibase 10.1016/0039-6028(94)90681-5} {\bibfield
  {journal} {\bibinfo  {journal} {Surface Science}\ }\textbf {\bibinfo {volume}
  {299-300}},\ \bibinfo {pages} {551} (\bibinfo {year} {1994})}\BibitemShut
  {NoStop}%
\bibitem [{\citenamefont {Yamada}\ and\ \citenamefont
  {Kimura}(1994)}]{Yamada1994}%
  \BibitemOpen
  \bibfield  {author} {\bibinfo {author} {\bibfnamefont {C.}~\bibnamefont
  {Yamada}}\ and\ \bibinfo {author} {\bibfnamefont {T.}~\bibnamefont
  {Kimura}},\ }\href {\doibase 10.1103/PhysRevB.49.14372} {\bibfield  {journal}
  {\bibinfo  {journal} {Physical Review B}\ }\textbf {\bibinfo {volume} {49}},\
  \bibinfo {pages} {14372} (\bibinfo {year} {1994})}\BibitemShut {NoStop}%
\bibitem [{\citenamefont {Jordan}\ \emph {et~al.}(1997)\citenamefont {Jordan},
  \citenamefont {Schillinger}, \citenamefont {Dressler}, \citenamefont
  {Karmann}, \citenamefont {Richter}, \citenamefont {Goetz}, \citenamefont
  {Marowsky},\ and\ \citenamefont {Sauerbrey}}]{Jordan1997}%
  \BibitemOpen
  \bibfield  {author} {\bibinfo {author} {\bibfnamefont {C.}~\bibnamefont
  {Jordan}}, \bibinfo {author} {\bibfnamefont {H.}~\bibnamefont {Schillinger}},
  \bibinfo {author} {\bibfnamefont {L.}~\bibnamefont {Dressler}}, \bibinfo
  {author} {\bibfnamefont {S.}~\bibnamefont {Karmann}}, \bibinfo {author}
  {\bibfnamefont {W.}~\bibnamefont {Richter}}, \bibinfo {author} {\bibfnamefont
  {K.}~\bibnamefont {Goetz}}, \bibinfo {author} {\bibfnamefont
  {G.}~\bibnamefont {Marowsky}}, \ and\ \bibinfo {author} {\bibfnamefont
  {R.}~\bibnamefont {Sauerbrey}},\ }\href {\doibase 10.1007/s003390050574}
  {\bibfield  {journal} {\bibinfo  {journal} {Applied Physics A: Materials
  Science {\&} Processing}\ }\textbf {\bibinfo {volume} {65}},\ \bibinfo
  {pages} {251} (\bibinfo {year} {1997})}\BibitemShut {NoStop}%
\bibitem [{\citenamefont {Niedermeier}\ \emph {et~al.}(1999)\citenamefont
  {Niedermeier}, \citenamefont {Schillinger}, \citenamefont {Sauerbrey},
  \citenamefont {Adolph},\ and\ \citenamefont {Bechstedt}}]{Niedermeier1999}%
  \BibitemOpen
  \bibfield  {author} {\bibinfo {author} {\bibfnamefont {S.}~\bibnamefont
  {Niedermeier}}, \bibinfo {author} {\bibfnamefont {H.}~\bibnamefont
  {Schillinger}}, \bibinfo {author} {\bibfnamefont {R.}~\bibnamefont
  {Sauerbrey}}, \bibinfo {author} {\bibfnamefont {B.}~\bibnamefont {Adolph}}, \
  and\ \bibinfo {author} {\bibfnamefont {F.}~\bibnamefont {Bechstedt}},\ }\href
  {\doibase 10.1063/1.124459} {\bibfield  {journal} {\bibinfo  {journal}
  {Applied Physics Letters}\ }\textbf {\bibinfo {volume} {75}},\ \bibinfo
  {pages} {618} (\bibinfo {year} {1999})}\BibitemShut {NoStop}%
\bibitem [{\citenamefont {Dekorsy}\ \emph {et~al.}(2003)\citenamefont
  {Dekorsy}, \citenamefont {Yakovlev}, \citenamefont {Seidel}, \citenamefont
  {Helm},\ and\ \citenamefont {Keilmann}}]{Dekorsy2003}%
  \BibitemOpen
  \bibfield  {author} {\bibinfo {author} {\bibfnamefont {T.}~\bibnamefont
  {Dekorsy}}, \bibinfo {author} {\bibfnamefont {V.}~\bibnamefont {Yakovlev}},
  \bibinfo {author} {\bibfnamefont {W.}~\bibnamefont {Seidel}}, \bibinfo
  {author} {\bibfnamefont {M.}~\bibnamefont {Helm}}, \ and\ \bibinfo {author}
  {\bibfnamefont {F.}~\bibnamefont {Keilmann}},\ }\href {\doibase
  10.1103/PhysRevLett.90.055508} {\bibfield  {journal} {\bibinfo  {journal}
  {Physical Review Letters}\ }\textbf {\bibinfo {volume} {90}},\ \bibinfo
  {pages} {055508} (\bibinfo {year} {2003})}\BibitemShut {NoStop}%
\bibitem [{\citenamefont {Bovino}\ \emph {et~al.}(2013)\citenamefont {Bovino},
  \citenamefont {Tasco}, \citenamefont {Passaseo}, \citenamefont {Larciprete},
  \citenamefont {Belardini},\ and\ \citenamefont {Sibilia}}]{Bovino2013}%
  \BibitemOpen
  \bibfield  {author} {\bibinfo {author} {\bibfnamefont {F.~A.}\ \bibnamefont
  {Bovino}}, \bibinfo {author} {\bibfnamefont {V.}~\bibnamefont {Tasco}},
  \bibinfo {author} {\bibfnamefont {A.}~\bibnamefont {Passaseo}}, \bibinfo
  {author} {\bibfnamefont {M.~C.}\ \bibnamefont {Larciprete}}, \bibinfo
  {author} {\bibfnamefont {A.}~\bibnamefont {Belardini}}, \ and\ \bibinfo
  {author} {\bibfnamefont {C.}~\bibnamefont {Sibilia}},\ }\href {\doibase
  10.1364/JOSAB.31.000026} {\bibfield  {journal} {\bibinfo  {journal} {Journal
  of the Optical Society of America B}\ }\textbf {\bibinfo {volume} {31}},\
  \bibinfo {pages} {26} (\bibinfo {year} {2013})}\BibitemShut {NoStop}%
\bibitem [{\citenamefont {Fiebig}\ \emph {et~al.}(2005)\citenamefont {Fiebig},
  \citenamefont {Pavlov},\ and\ \citenamefont {Pisarev}}]{Fiebig2005}%
  \BibitemOpen
  \bibfield  {author} {\bibinfo {author} {\bibfnamefont {M.}~\bibnamefont
  {Fiebig}}, \bibinfo {author} {\bibfnamefont {V.~V.}\ \bibnamefont {Pavlov}},
  \ and\ \bibinfo {author} {\bibfnamefont {R.~V.}\ \bibnamefont {Pisarev}},\
  }\href {\doibase 10.1364/JOSAB.22.000096} {\bibfield  {journal} {\bibinfo
  {journal} {Journal of the Optical Society of America B}\ }\textbf {\bibinfo
  {volume} {22}},\ \bibinfo {pages} {96} (\bibinfo {year} {2005})}\BibitemShut
  {NoStop}%
\bibitem [{\citenamefont {Becher}\ \emph {et~al.}(2015)\citenamefont {Becher},
  \citenamefont {Maurel}, \citenamefont {Aschauer}, \citenamefont {Lilienblum},
  \citenamefont {Mag{\'{e}}n}, \citenamefont {Meier}, \citenamefont
  {Langenberg}, \citenamefont {Trassin}, \citenamefont {Blasco}, \citenamefont
  {Krug}, \citenamefont {Algarabel}, \citenamefont {Spaldin}, \citenamefont
  {Pardo},\ and\ \citenamefont {Fiebig}}]{Becher2015}%
  \BibitemOpen
  \bibfield  {author} {\bibinfo {author} {\bibfnamefont {C.}~\bibnamefont
  {Becher}}, \bibinfo {author} {\bibfnamefont {L.}~\bibnamefont {Maurel}},
  \bibinfo {author} {\bibfnamefont {U.}~\bibnamefont {Aschauer}}, \bibinfo
  {author} {\bibfnamefont {M.}~\bibnamefont {Lilienblum}}, \bibinfo {author}
  {\bibfnamefont {C.}~\bibnamefont {Mag{\'{e}}n}}, \bibinfo {author}
  {\bibfnamefont {D.}~\bibnamefont {Meier}}, \bibinfo {author} {\bibfnamefont
  {E.}~\bibnamefont {Langenberg}}, \bibinfo {author} {\bibfnamefont
  {M.}~\bibnamefont {Trassin}}, \bibinfo {author} {\bibfnamefont
  {J.}~\bibnamefont {Blasco}}, \bibinfo {author} {\bibfnamefont {I.~P.}\
  \bibnamefont {Krug}}, \bibinfo {author} {\bibfnamefont {P.~a.}\ \bibnamefont
  {Algarabel}}, \bibinfo {author} {\bibfnamefont {N.~a.}\ \bibnamefont
  {Spaldin}}, \bibinfo {author} {\bibfnamefont {J.~a.}\ \bibnamefont {Pardo}},
  \ and\ \bibinfo {author} {\bibfnamefont {M.}~\bibnamefont {Fiebig}},\ }\href
  {\doibase 10.1038/nnano.2015.108} {\bibfield  {journal} {\bibinfo  {journal}
  {Nature Nanotechnology}\ }\textbf {\bibinfo {volume} {10}},\ \bibinfo {pages}
  {661} (\bibinfo {year} {2015})}\BibitemShut {NoStop}%
\bibitem [{\citenamefont {Li}\ \emph {et~al.}(2013)\citenamefont {Li},
  \citenamefont {Rao}, \citenamefont {Mak}, \citenamefont {You}, \citenamefont
  {Wang}, \citenamefont {Dean},\ and\ \citenamefont {Heinz}}]{Li2013a}%
  \BibitemOpen
  \bibfield  {author} {\bibinfo {author} {\bibfnamefont {Y.}~\bibnamefont
  {Li}}, \bibinfo {author} {\bibfnamefont {Y.}~\bibnamefont {Rao}}, \bibinfo
  {author} {\bibfnamefont {K.~F.}\ \bibnamefont {Mak}}, \bibinfo {author}
  {\bibfnamefont {Y.}~\bibnamefont {You}}, \bibinfo {author} {\bibfnamefont
  {S.}~\bibnamefont {Wang}}, \bibinfo {author} {\bibfnamefont {C.~R.}\
  \bibnamefont {Dean}}, \ and\ \bibinfo {author} {\bibfnamefont {T.~F.}\
  \bibnamefont {Heinz}},\ }\href {\doibase 10.1021/nl401561r} {\bibfield
  {journal} {\bibinfo  {journal} {Nano letters}\ }\textbf {\bibinfo {volume}
  {13}},\ \bibinfo {pages} {3329} (\bibinfo {year} {2013})}\BibitemShut
  {NoStop}%
\bibitem [{\citenamefont {Zhao}\ \emph {et~al.}(2015)\citenamefont {Zhao},
  \citenamefont {Torchinsky}, \citenamefont {Chu}, \citenamefont {Ivanov},
  \citenamefont {Lifshitz}, \citenamefont {Flint}, \citenamefont {Qi},
  \citenamefont {Cao},\ and\ \citenamefont {Hsieh}}]{Zhao2015}%
  \BibitemOpen
  \bibfield  {author} {\bibinfo {author} {\bibfnamefont {L.}~\bibnamefont
  {Zhao}}, \bibinfo {author} {\bibfnamefont {D.~H.}\ \bibnamefont
  {Torchinsky}}, \bibinfo {author} {\bibfnamefont {H.}~\bibnamefont {Chu}},
  \bibinfo {author} {\bibfnamefont {V.}~\bibnamefont {Ivanov}}, \bibinfo
  {author} {\bibfnamefont {R.}~\bibnamefont {Lifshitz}}, \bibinfo {author}
  {\bibfnamefont {R.}~\bibnamefont {Flint}}, \bibinfo {author} {\bibfnamefont
  {T.}~\bibnamefont {Qi}}, \bibinfo {author} {\bibfnamefont {G.}~\bibnamefont
  {Cao}}, \ and\ \bibinfo {author} {\bibfnamefont {D.}~\bibnamefont {Hsieh}},\
  }\href {\doibase 10.1038/nphys3517} {\bibfield  {journal} {\bibinfo
  {journal} {Nature Physics}\ }\textbf {\bibinfo {volume} {12}},\ \bibinfo
  {pages} {32} (\bibinfo {year} {2015})}\BibitemShut {NoStop}%
\bibitem [{\citenamefont {Paarmann}\ \emph {et~al.}(2015)\citenamefont
  {Paarmann}, \citenamefont {Razdolski}, \citenamefont {Melnikov},
  \citenamefont {Gewinner}, \citenamefont {Sch{\"{o}}llkopf},\ and\
  \citenamefont {Wolf}}]{Paarmann2015}%
  \BibitemOpen
  \bibfield  {author} {\bibinfo {author} {\bibfnamefont {A.}~\bibnamefont
  {Paarmann}}, \bibinfo {author} {\bibfnamefont {I.}~\bibnamefont {Razdolski}},
  \bibinfo {author} {\bibfnamefont {A.}~\bibnamefont {Melnikov}}, \bibinfo
  {author} {\bibfnamefont {S.}~\bibnamefont {Gewinner}}, \bibinfo {author}
  {\bibfnamefont {W.}~\bibnamefont {Sch{\"{o}}llkopf}}, \ and\ \bibinfo
  {author} {\bibfnamefont {M.}~\bibnamefont {Wolf}},\ }\href {\doibase
  10.1063/1.4929358} {\bibfield  {journal} {\bibinfo  {journal} {Applied
  Physics Letters}\ }\textbf {\bibinfo {volume} {107}},\ \bibinfo {pages}
  {081101} (\bibinfo {year} {2015})}\BibitemShut {NoStop}%
\bibitem [{\citenamefont {Sch{\"{o}}llkopf}\ \emph {et~al.}(2015)\citenamefont
  {Sch{\"{o}}llkopf}, \citenamefont {Gewinner}, \citenamefont {Junkes},
  \citenamefont {Paarmann}, \citenamefont {von Helden}, \citenamefont {Bluem},\
  and\ \citenamefont {Todd}}]{Schollkopf2015}%
  \BibitemOpen
  \bibfield  {author} {\bibinfo {author} {\bibfnamefont {W.}~\bibnamefont
  {Sch{\"{o}}llkopf}}, \bibinfo {author} {\bibfnamefont {S.}~\bibnamefont
  {Gewinner}}, \bibinfo {author} {\bibfnamefont {H.}~\bibnamefont {Junkes}},
  \bibinfo {author} {\bibfnamefont {A.}~\bibnamefont {Paarmann}}, \bibinfo
  {author} {\bibfnamefont {G.}~\bibnamefont {von Helden}}, \bibinfo {author}
  {\bibfnamefont {H.}~\bibnamefont {Bluem}}, \ and\ \bibinfo {author}
  {\bibfnamefont {A.~M.~M.}\ \bibnamefont {Todd}},\ }\href {\doibase
  10.1117/12.2182284} {\bibfield  {journal} {\bibinfo  {journal} {Proc. SPIE}\
  }\textbf {\bibinfo {volume} {9512}},\ \bibinfo {pages} {95121L} (\bibinfo
  {year} {2015})}\BibitemShut {NoStop}%
\bibitem [{\citenamefont {{Y. R. Shen}}(1989)}]{Shen1989}%
  \BibitemOpen
  \bibfield  {author} {\bibinfo {author} {\bibnamefont {{Y. R. Shen}}},\ }\href
  {\doibase 10.1146/annurev.physchem.40.1.327} {\bibfield  {journal} {\bibinfo
  {journal} {Annual Review of Physical Chemistry}\ }\textbf {\bibinfo {volume}
  {40}},\ \bibinfo {pages} {327} (\bibinfo {year} {1989})}\BibitemShut
  {NoStop}%
\bibitem [{\citenamefont {Shen}(2003)}]{Shen2003}%
  \BibitemOpen
  \bibfield  {author} {\bibinfo {author} {\bibfnamefont {Y.~R.}\ \bibnamefont
  {Shen}},\ }\href@noop {} {\emph {\bibinfo {title} {{The Principles of
  Nonlinear Optics}}}}\ (\bibinfo  {publisher} {Wiley-Interscience},\ \bibinfo
  {year} {2003})\BibitemShut {NoStop}%
\bibitem [{\citenamefont {Flytzanis}(1972)}]{Flytzanis1972}%
  \BibitemOpen
  \bibfield  {author} {\bibinfo {author} {\bibfnamefont {C.}~\bibnamefont
  {Flytzanis}},\ }\href {\doibase 10.1103/PhysRevB.6.1264} {\bibfield
  {journal} {\bibinfo  {journal} {Physical Review B}\ }\textbf {\bibinfo
  {volume} {6}},\ \bibinfo {pages} {1264} (\bibinfo {year} {1972})}\BibitemShut
  {NoStop}%
\bibitem [{\citenamefont {Faust}\ and\ \citenamefont
  {Henry}(1966)}]{Faust1966}%
  \BibitemOpen
  \bibfield  {author} {\bibinfo {author} {\bibfnamefont {W.}~\bibnamefont
  {Faust}}\ and\ \bibinfo {author} {\bibfnamefont {C.}~\bibnamefont {Henry}},\
  }\href {\doibase 10.1103/PhysRevLett.17.1265} {\bibfield  {journal} {\bibinfo
   {journal} {Physical Review Letters}\ }\textbf {\bibinfo {volume} {17}},\
  \bibinfo {pages} {1265} (\bibinfo {year} {1966})}\BibitemShut {NoStop}%
\bibitem [{\citenamefont {Roman}\ \emph {et~al.}(2006)\citenamefont {Roman},
  \citenamefont {Yates}, \citenamefont {Veithen}, \citenamefont {Vanderbilt},\
  and\ \citenamefont {Souza}}]{Roman2006}%
  \BibitemOpen
  \bibfield  {author} {\bibinfo {author} {\bibfnamefont {E.}~\bibnamefont
  {Roman}}, \bibinfo {author} {\bibfnamefont {J.}~\bibnamefont {Yates}},
  \bibinfo {author} {\bibfnamefont {M.}~\bibnamefont {Veithen}}, \bibinfo
  {author} {\bibfnamefont {D.}~\bibnamefont {Vanderbilt}}, \ and\ \bibinfo
  {author} {\bibfnamefont {I.}~\bibnamefont {Souza}},\ }\href {\doibase
  10.1103/PhysRevB.74.245204} {\bibfield  {journal} {\bibinfo  {journal}
  {Physical Review B}\ }\textbf {\bibinfo {volume} {74}},\ \bibinfo {pages}
  {245204} (\bibinfo {year} {2006})}\BibitemShut {NoStop}%
\bibitem [{\citenamefont {{Mosteller, Jr.}}\ and\ \citenamefont
  {Wooten}(1968)}]{Mosteller1968}%
  \BibitemOpen
  \bibfield  {author} {\bibinfo {author} {\bibfnamefont {L.~P.}\ \bibnamefont
  {{Mosteller, Jr.}}}\ and\ \bibinfo {author} {\bibfnamefont {F.}~\bibnamefont
  {Wooten}},\ }\href {\doibase 10.1364/JOSA.58.000511} {\bibfield  {journal}
  {\bibinfo  {journal} {Journal of the Optical Society of America}\ }\textbf
  {\bibinfo {volume} {58}},\ \bibinfo {pages} {511} (\bibinfo {year}
  {1968})}\BibitemShut {NoStop}%
\bibitem [{\citenamefont {Lekner}(1991)}]{Lekner1991}%
  \BibitemOpen
  \bibfield  {author} {\bibinfo {author} {\bibfnamefont {J.}~\bibnamefont
  {Lekner}},\ }\href {\doibase 10.1088/0953-8984/3/32/017} {\bibfield
  {journal} {\bibinfo  {journal} {Journal of Physics: Condensed Matter}\
  }\textbf {\bibinfo {volume} {3}},\ \bibinfo {pages} {6121} (\bibinfo {year}
  {1991})}\BibitemShut {NoStop}%
\bibitem [{\citenamefont {Razdolski}\ \emph {et~al.}(2016)\citenamefont
  {Razdolski}, \citenamefont {Chen}, \citenamefont {Giles}, \citenamefont
  {Gewinner}, \citenamefont {Sch{\"{o}}llkopf}, \citenamefont {Minghui},
  \citenamefont {Wolf}, \citenamefont {Giannini}, \citenamefont {Caldwell},
  \citenamefont {Maier},\ and\ \citenamefont {Paarmann}}]{Razdolski2016}%
  \BibitemOpen
  \bibfield  {author} {\bibinfo {author} {\bibfnamefont {I.}~\bibnamefont
  {Razdolski}}, \bibinfo {author} {\bibfnamefont {Y.}~\bibnamefont {Chen}},
  \bibinfo {author} {\bibfnamefont {A.~J.}\ \bibnamefont {Giles}}, \bibinfo
  {author} {\bibfnamefont {S.}~\bibnamefont {Gewinner}}, \bibinfo {author}
  {\bibfnamefont {W.}~\bibnamefont {Sch{\"{o}}llkopf}}, \bibinfo {author}
  {\bibfnamefont {M.~H.}~\bibnamefont {Hong}}, \bibinfo {author} {\bibfnamefont
  {M.}~\bibnamefont {Wolf}}, \bibinfo {author} {\bibfnamefont {V.}~\bibnamefont
  {Giannini}}, \bibinfo {author} {\bibfnamefont {J.~D.}\ \bibnamefont
  {Caldwell}}, \bibinfo {author} {\bibfnamefont {S.~A.}\ \bibnamefont {Maier}},
  \ and\ \bibinfo {author} {\bibfnamefont {A.}~\bibnamefont {Paarmann}},\
  }
  \Eprint {http://arxiv.org/abs/1607.05158}
  {arXiv:1607.05158} (\bibinfo {year} {2016})
   \BibitemShut {NoStop}%
\end{thebibliography}
%

\end{document}